# Micrometer-scale Magnetic Imaging of Geological Samples Using a Quantum Diamond Microscope

D. R. Glenn[1], R. R. Fu[2], P. Kehayias[3], D. Le Sage[1], E. A. Lima[2], B. P. Weiss[2], R. L. Walsworth[1,3]

1) Department of Physics, Harvard University, Cambridge, Massachusetts, USA.
2) Department of Earth, Atmospheric, and Planetary Sciences, Massachusetts Institute of Technology, Cambridge, Massachusetts, USA.
3) Harvard-Smithsonian Center for Astrophysics, Cambridge, Massachusetts, USA.

Corresponding author: Ronald Walsworth (rwalsworth@cfa.harvard.edu)

**Key Points:**

- We developed a quantum diamond microscope (QDM) for imaging magnetic fields produced by geological samples.
- This instrument offers a combination of superior spatial resolution (5 µm), magnetic sensitivity (20 µT· µm/Hz1/2), and wide field of view (4 mm).
- We demonstrate the application of this instrument to magnetic mapping of several terrestrial and meteoritic rock samples.

# Abstract

Remanent magnetization in geological samples may record the past intensity and direction of planetary magnetic fields. Traditionally, this magnetization is analyzed through measurements of the net magnetic moment of bulk millimeter to centimeter sized samples. However, geological samples are often mineralogically and texturally heterogeneous at submillimeter scales, with only a fraction of the ferromagnetic grains carrying the remanent magnetization of interest. Therefore, characterizing this magnetization in such cases requires a technique capable of imaging magnetic fields at fine spatial scales and with high sensitivity. To address this challenge, we developed a new instrument, based on nitrogen-vacancy centers in diamond, which enables direct imaging of magnetic fields due to both remanent and induced magnetization, as well as optical imaging, of room-temperature geological samples with spatial resolution approaching the optical diffraction limit. We describe the operating principles of this device, which we call the quantum diamond microscope (QDM), and report its optimized image-area-normalized magnetic field sensitivity (20 µT·µm/Hz½), spatial resolution (5 µm), and field of view (4 mm), as well as trade-offs between these parameters. We also perform an absolute magnetic field calibration for the device in different modes of operation, including three-axis (vector) and single-axis (projective) magnetic field imaging. Finally, we use the QDM to obtain magnetic images of several terrestrial and meteoritic rock samples, demonstrating its ability to resolve spatially distinct populations of ferromagnetic carriers.

# Introduction

Reliable characterization of ancient magnetic fields using geological samples requires the survival of ferromagnetic phases that were present to record the magnetic field of interest. However, metamorphism, aqueous alteration, and surface weathering are common processes that may destroy and replace all or a subset of a rock sample's ferromagnetic minerals. At the same time, erosion may lead to the physical disintegration of rocks and the reassembly of their constituent parts in clastic sediments. As a result of these secondary chemical and physical processes, diachronous populations of ferromagnetic minerals are frequently juxtaposed at the micrometer to millimeter scales. The difficulty of extracting useful paleomagnetic information from such complex samples is a key limiting factor to our understanding of the early Earth magnetic field (e.g., [Weiss *et. al.* 2015]), the past configuration of lithospheric plates (e.g., [Kent *et. al.* 1987]), and the stability of the Earth's rotation and geomagnetic field axis in the mantle reference frame (e.g., [Swanson-Hysell *et. al.* 2012]). Furthermore, even in the absence of secondary remagnetization, geological samples often contain different populations of ferromagnetic minerals with disparate capacities for recording a high-fidelity, interpretable remanence [Fu *et. al.* 2014].

Obtaining useful paleomagnetic data from samples with complex geological histories and/or fine-scale mixtures of ferromagnetic grains with varying recording properties requires the isolation of magnetic signal carried by high-fidelity ferromagnetic grains that were magnetized during the time period of interest. Standard paleomagnetic techniques, limited in part by the magnetometer sensitivity, typically measure the net magnetic moment of whole-rock samples with volumes of several cubic centimeters. For such samples, magnetic cleaning techniques such as thermal and alternating field demagnetization have been used to isolate the magnetization carried by sub-populations of ferromagnetic grains [Tauxe 2010]. However, magnetic cleaning often does not fully isolate a single population of grains due to overlapping ranges of unblocking temperatures or coercivities. Furthermore, because the microscopic contexts of ferromagnetic grain populations are generally not directly constrained by magnetic cleaning, correctly identifying the relative age and origin of the magnetization component of interest is often ambiguous.

The development of mapping magnetometers capable of resolving ferromagnetic carriers at the submillimeter scale — such as the SQUID microscope [Weiss 2007 et. al., Kawai et. al. 2016], magneto-impedance microscopes [Nakamura et. al. 2010], magnetoresistance (MR) microscopes [Hankard et. al. 2009], magneto-optical imaging (MOI) [Uehara et. al. 2010], Hall-effect microscopes [Kletetschka et. al. 2013] and the magnetic tunnel junction (MTJ) microscope [Lima et. al. 2014] — has led to a new approach for analyzing rock samples with complex, heterogeneous magnetizations. These instruments are capable of imaging the spatial distribution of remanent fields so that they can be spatially correlated with the fine-scale magnetization distribution within a sample [Fu et. al. 2014].

It has recently been demonstrated that nitrogen-vacancy (NV) centers in diamond [Doherty et. al. 2013, Schirhagl et. al. 2014, Rondin et. al. 2014] enable sensitive imaging of static magnetic field distributions with sub-micrometer spatial resolution [Pham et. al. 2011]. Perhaps the simplest and most robust implementation of NV magnetic imaging is the quantum diamond microscope (QDM) [Glenn et. al. 2015]. This device consists of a thin (10 nm – 10 µm) layer of NV centers implanted or grown into the surface of a diamond chip that is brought into close proximity with the magnetic sources of interest and interrogated using optically detected magnetic resonance (ODMR) [Gruber et. al. 1997]. The QDM is readily applicable to the study of geological samples at room temperature (< 50° C) and low ambient magnetic field (< 50 µT), making it an attractive platform for spatially-resolved measurements of artificial and natural remanent magnetization (NRM). Furthermore, because the NV centers can typically be placed within ~10 µm of the magnetic sources of interest (or closer, if special care is taken with sample and sensor preparation), even a relatively modest QDM magnetic noise floor of ~100 nT RMS is sufficient to detect moments as small as ~$10^{-16}$ A·m$^2$, providing sensitivity comparable or superior to scanning SQUID, MR, MOI, Hall-effect, and MTJ microscope technologies. The QDM also allows optical imaging in the same configuration, providing spatially-correlated magnetic and optical images of the sample.

## 2. Description of the QDM

### 2.1 Basic Operating Principles

In the QDM implementation described here (Figure 1a-b), a diamond sensor chip (labeled D1 – D4 in our experiments; see Supplementary Table S1 for details) is placed, NV layer face down, above the polished surface of the rock sample. The sample is mounted on a printed circuit board patterned with a pair of decoupled, crossed stripline microwave resonators. The striplines are excited in-phase (or 90° out of phase) to produce a linearly- (or circularly-) polarized, GHz-frequency magnetic field for driving continuous spin rotations in the NV centers. The NV centers are continuously probed with green laser excitation (wavelength 532 nm, intensity $10^5 – 10^7$ W/m$^2$), and their emitted red fluorescence (wavelength 637 – 800 nm) is imaged with a scientific complementary metal oxide semiconductor (sCMOS) camera.

For geological samples that could be adversely affected by heating during prolonged laser illumination, the diamond chip may be angle-polished such that excitation light is totally internally reflected from the bottom surface and exits the opposite facet (Figure 1b). In contrast, with a square-cut chip and laser light directly impinging on the sample, we observe up to ~30 °C heating above room temperature under realistic imaging conditions for most sample substrates (Supplementary Figure S2).

Three pairs of Helmholtz-configured magnet coils produce a spatially homogeneous static field ($B_0$) at the rock sample and NV layer, which can be used to cancel the local Earth's field and provide the NV centers with a controlled magnetic bias. The accuracy of the field nulling achieved with the present system is ~0.1 µT, comparable to the magnetic sensitivity of our measurements, but could be improved with the use of high-precision current supplies and/or secondary shim coils if needed.

The ODMR measurement proceeds by simultaneous application of optical and microwave fields (the latter denoted $B_1$) to the ensemble of NV centers, generating a spatially-dependent NV fluorescence signal with features dependent on the sample-induced, local magnetic field. (See Supplementary Figure S3 for background on the physics and level structure of NV centers in diamond.) In the absence of microwaves, the optical excitation induces a baseline fluorescence level by continuously driving the NV population into an excited electronic state, which spontaneously decays (lifetime $\tau \approx 13$ ns [Robledo *et. al.* 2011]) by emitting red fluorescence. The addition of a microwave B1 field resonant with one of the ground state spin transitions causes a fraction of the NV population to be transferred from the $m_s$ = 0 to the $m_s$ = -1 or $m_s$ = +1 states. These states, when optically excited, have a finite branching ratio to a long-lived metastable state (lifetime $\tau \approx 220$ ns [Acosta *et. al.* 2010]), decreasing the emitted fluorescence. Off-resonant microwave driving leaves the baseline fluorescence unchanged. The spin resonance frequency for each NV center is Zeeman-shifted by the local magnetic field, resulting in a spatially-varying fluorescence pattern where the light emitted at each position is suppressed below the baseline only if the microwave drive is resonant. In the low-field limit ($g\,\mu_B\,B \ll f_{ZFS}$, for Landé g-factor $g \approx 2$, $\mu_B \approx 14$ GHz/T the Bohr magneton, and $f_{ZFS} \approx 2.87$ GHz the NV zero-field splitting), the Zeeman shift is $\Delta f = \pm\,g\,\mu_B\,|B_p|$, where $B_p$ is the magnitude of the local magnetic field projected onto the NV axis. The sign of the shift is positive (negative) for the $\Delta m_s$ =+1 ($\Delta m_s$ =-1) spin transition. By sweeping the microwave frequency and collecting a fluorescence image at each increment of microwave frequency, the magnetic field spatial distribution is obtained (Supplementary Figure S4). The resulting field map may then be spatially filtered in software to optimize sensitivity at length scales of interest (typically ~5 – 500 µm for the geological samples investigated in this work).

To correlate magnetic field maps with mineralogical properties of the sample, reflected-light microscopy is carried out in-place in the QDM. The green laser excitation beam is turned off and replaced with light from a red light-emitting diode (LED, wavelength ~660 nm). The red light passes through the diamond, reflects off the surface of the sample, and is imaged onto the camera using the same collection optics as for NV fluorescence. In the QDM system we describe here, spherical aberration due to imaging through the diamond chip (~500 µm thickness) limits the spatial resolution of reflected-light microscopy to ~5 µm. This is comparable to the spatial resolution of magnetic imaging in the present QDM system (also limited by spherical aberration, as well as sample roughness, which is of similar magnitude), and sufficient in practice to enable co-registration with high resolution optical images acquired in other instruments. The problem of spherical aberration in the QDM can be overcome by designing thinner diamond chips, or by introducing additional correcting lenses into the detection path.

We note that optical diffraction fundamentally limits spatial resolution, in both magnetic maps and reflected-light microscopy. Using a high numerical aperture (NA) objective lens, the diffraction limit is about ~0.4 µm for our camera-based QDM implementation [LeSage *et. al.* 2013]. Diffraction can, in principle, be circumvented by using super-resolution imaging techniques with scanning excitation [Maurer *et. al.* 2010], making the QDM potentially suitable for direct imaging of single ferromagnetic grains in sub-micrometer thickness sections.

## 2.2 Modes of QDM Operation

### 2.2.1 Vector Magnetic Microscopy (VMM)

The choice of $B_0$ and $B_1$ allows the QDM instrument to be operated in several different modes, with associated tradeoffs in sensitivity, ease of calibration, and ambient field cancellation. The most general approach is NV vector magnetic microscopy (VMM), in which the $B_0$ field is aligned with nonzero, unequal projections onto all four NV orientations (each coinciding with one of the four [111] diamond crystal axes). The magnitude of $B_0$ is typically set to ~1 mT such that all ODMR lines are fully resolved (Figure 1c). The frequency of the linearly-polarized $B_1$ field is swept across the entire range of NV resonances, providing full (three-axis) vector magnetic field information in each pixel of the image. The current applied to the Helmholtz coils is then reversed and the measurement repeated. The sum of these two measured field maps represents the signal from remanent magnetization carried by ferromagnetic grains with coercivity > $B_0$. The difference between maps represents the induced magnetization, which may be used to localize paramagnetic, superparamagnetic, and other low-coercivity carriers. We achieve a field reversal precision of ~$10^{-4}$ in the present system, corresponding to a residual bias $B_0^{(r)}$ < 0.1 µT in the ferromagnetic (measurement sum) field map. (The residual bias $B_0^{(r)}$ is defined as the vector sum of the positively- and negatively- oriented applications of B0, and is measured in a region of the field map far from any magnetic sources.) Using the VMM technique, we obtain an image-area-normalized sensitivity of about ~1 – 100 µT·µm/Hz$^{½}$, depending on details of the NV-diamond sensor and the configuration of fluorescence excitation and collection optics.

### 2.2.2 Projective Magnetic Microscopy (PMM)

In the second ODMR technique, we restrict the set of microwave transition frequencies scanned to enable faster data acquisition. In general, magnetic field sensitivity is improved by reducing microwave scan range, which is feasible if all regions of the sample produce magnetic fields that lie in a restricted region of the Zeeman-shifted ODMR spectrum. For applications that require only single-axis magnetic imaging, $B_0$ may be aligned parallel to just one of the [111] diamond lattice directions, such that the spin transition frequencies for the NV orientation along that orientation are maximally split. Transition frequencies for the other three orientations are degenerate due to tetrahedral symmetry of the lattice.

The ODMR frequency $f$ is swept only over the selected pair of resonances, and ideally over only over the part of each resonance with maximum slope, d$S$/d$f$, for $S$ the ODMR fluorescence signal contrast. This can result in particularly efficient sensing due to the reduced signal acquisition time per sweep, and the strong dependence of the fluorescence on small changes in the local magnetic field (Figure 1d). In addition, using only one NV orientation provides better ODMR contrast by allowing better optimization of the green light polarization. We refer to this approach as projective magnetic microscopy (PMM), because, to a good approximation (Supplementary Figure S4), only the component of the sample field projected along the selected NV orientation is detected. As with VMM, each set of PMM measurements consists of magnetic field maps acquired with an aligned and an anti-aligned bias field relative to the chosen [111] lattice direction. The sum and difference of these two maps yield the magnetic fields due to remanent and induced magnetizations, respectively. PMM can enhance sensitivity by a factor of ~2 – 3 compared with typical VMM, although the largest gains are available only if the range of sample magnetic field values to be imaged, $\Delta B_{samp}$, is small, (i.e., $\mu_B \Delta B_{samp} \ll \Gamma_{NV}$, for $\Gamma_{NV}$ the NV ODMR half width and at half maximum (HWHM) linewidth defined in Supplementary Figure 4), such that the frequency sweep range may be restricted to the sharpest part of the resonance [Glenn *et. al.* 2015]. This tradeoff between sensitivity and dynamic range is generic for NV ODMR measurements. Another potential drawback of PMM is that CVD diamond chips are typically grown with [100] normal to the sensing plane, such magnetic projection direction in PMM measurements is at an angle of 54.7° to the normal. Diamond chips with [111] normal to the plane can be produced by cleaving or polishing at the appropriate angle, usually at the expense of a smaller available field of view.

### 2.2.3 Circularly-Polarized Magnetic Microscopy (CPMM)

Bias fields $B_0 \geq 1$ mT are necessary during data acquisition in the VMM and PMM configurations to spectrally resolve the projected field magnitude along different NV orientations. Although the sum of successive measurements under a reversed $B_0$ rejects the induced magnetic response of the sample to $B_0$ during a measurement, down to the residual bias $B_0^{(r)} < 0.1$ µT, inaccuracy in the reversal of the bias field may introduce systematic errors. To avoid this potential problem and recover only the magnetic fields due to remanent magnetization without exposing the sample to significant magnetic fields, we adapt a single-axis ODMR technique (Figure 1e) previously demonstrated [Alegre *et. al.* 2007] to work at very small bias ($B_0 < 10$ µT in our system), by using a drive field $B_1$ that is circularly-polarized with respect to $B_0$.

The $B_0$ field is aligned normal to the diamond surface such that it projects equally on all NV orientations. In this configuration, the component of the sample magnetic field parallel to $B_0$ produces asymmetric shifts of the four sets of degenerate ODMR resonances relative to the zero-field splitting frequency, with the sign of the shift depending on the handedness of $B_1$. By switching between left- and right- circular drive fields, the centroid of the degenerate resonance lines is modulated to higher or lower frequency, allowing detection of small shifts, $g\,\mu_B\,B \ll \Gamma_{NV}$. The data shown in Figure 1e were obtained at $B_0 = 40$ µT, where the $\Delta m_s = +1$ and $\Delta m_s = -1$ transitions are resolved, to clearly illustrate the procedure. However, small line shifts can be detected even in the unresolved case, in practice. Sample magnetic fields perpendicular to $B_0$ give rise instead to a symmetric broadening of the degenerate resonances, and therefore cannot easily be detected. We refer to this approach as circularly-polarized magnetic microscopy (CPMM). CPMM is typically performed with NV centers formed from $^{15}$N, which have simpler spectra than NV centers formed from $^{14}$N (i.e. hyperfine doublets instead of triplets). Despite the extra technical overhead of the circularly-polarized NV drive, as well the need for careful calibration when strain terms in the NV Hamiltonian become comparable to the Zeeman term (see section 3.3 below), this is the only technique known to us that enables NV magnetic imaging at very low bias field. (See Supplementary Figure S5 for an extended discussion of the CPMM technique.)

## Measurements and Discussion

The high spatial resolution of the QDM, together with the ability to acquire optical and magnetic images simultaneously, enables local magnetic field maps to be precisely correlated with images of petrographic structure obtained in other instruments. To illustrate the constraints on ferromagnetic mineralogy offered by these measurements, we acquired co-registered optical and magnetic maps of three different rock samples, each of which displays heterogeneous magnetization at the 1-100 µm scale.

### 3.1 Correlative Optical and Magnetic Microscopy

As an example QDM application, we acquired spatially-correlated optical magnetic images of a 30 µm thin section of the Allende CV3 chondrite to identify the primary or secondary origin of ferromagnetic carriers. As in the case of most chondrites, Allende contains millimeter-scale igneous chondrules embedded in a finer-grain matrix consisting of primitive and recrystallized minerals. Early paleomagnetic studies have suggested that Allende chondrules recorded primordial solar system magnetic fields as they formed and cooled in the solar nebula [Lanoix *et. al.* 1978; Sugiura *et. al.* 1985]. However, subsequent paleomagnetic studies of Allende have revealed pervasive recrystallization of ferromagnetic phases, including the formation of pyrrhotite, magnetite, and awaruite during metasomatism on the CV parent body [Carporzen *et. al.* 2011; Fu *et. al.* 2014b]. The identification of ferromagnetic phases within individual chondrules is therefore necessary to assess the interpretation of chondrule remanent magnetization due to nebular magnetic fields. If the ferromagnetic mineralogy of chondrules consists of secondary minerals, the chondrules cannot preserve magnetization acquired prior to the accretion of the parent body.

To facilitate the identification of magnetic sources inside individual chondrules, we imparted a strong in-plane isothermal remanent magnetization (IRM) in a 200 mT field on a 30 µm thick section of Allende. Although this applied field does not saturate the pyrrhotite grains present in Allende, it covers the majority of the coercivity range within which the NRM is observed [Carporzen et al. 2011]. We therefore regard our maps of the 200 mT IRM as a satisfactory approximation of the distribution of ferromagnetic grains carrying NRM in the Allende meteorite. The diamond D1 (with $^{14}$NV layer depth of ~20 nm – see Supplementary Table S1) was placed directly onto the sample, such that the effective standoff distance to the sensor was set by the roughness of the polished sample surface (~1 – 3 µm, estimated by high-magnification optical microscopy). The full field of view (FOV) had dimensions ~800 µm × 600 µm. Comparison of the reflected light, plane-polarized image (Figure 2a) and VMM magnetic map (Figure 2b, acquired in $1.5 \times 10^3$ s of averaging) shows the concentration of ferromagnetic carriers in the rim and mesostasis regions of the chondrule. Reflected light microscopy indicates that the chondrule rim contains abundant magnetite. Meanwhile, although the ferromagnetic phases in the mesostasis regions of our measured chondrule were not visible in optical microscopy, previous analyses have shown that they also contain exclusively mineral assemblages formed during aqueous alteration [Brearley and Krot 2012]. As such, the spatial localization of ferromagnetism in these areas using the QDM indicates the secondary nature of ferromagnetic grains from Allende chondrules [Fu *et. al.* 2014b].

A similar application of the QDM to image a strong artificial magnetization has already been demonstrated in the literature [Fu *et. al.* 2014], as a means to identify the mineralogy of the strongest sources in the sample after the volume-averaged natural remanence had been extracted by conventional thermal or alternating-field demagnetization. In principle, it should be possible to obtain spatially-resolved demagnetization curves within each magnetically homogeneous sub-region of a QDM image directly, although this technology has yet to be implemented.

To test the repeatability of QDM magnetic field maps, as well as the agreement between different ODMR acquisition modes, we re-imaged a smaller field of view (sub-FOV) of the Allende chondrule containing several altered mesostasis regions using diamond D2 (with a $^{15}$NV layer depth of ~20 nm) in VMM and CPMM (Figure 2c). The diamond position and laser illumination pattern were kept fixed between the two measurements; only the direction of $B_0$ and the polarization and frequency sweep range of $B_1$ were changed. Each image was acquired in $2.0 \times 10^4$ s of averaging. After mean-subtraction to account for the difference in $B_0$, the field maps showed good qualitative agreement, with all of the dipolar magnetic features from the full FOV map clearly discernible in the new sub-FOVs. However, careful comparison shows that the peak field values of magnetic features measured in CPMM are greater than those in VMM, typically by ~5 – 10%. This suggests the need for calibration of the different ODMR imaging techniques, as described in the following sections. In addition, we note that the full FOV field map contains a prominent, sharp noise feature (white arrow) that does not appear in either of the sub-FOVs. We attribute this to local diamond-strain-induced shifts in NV resonance frequencies in the affected pixels, likely due to an edge dislocation in the diamond crystal running perpendicular to the chip surface. (Changing to a different diamond for the sub-FOVs thus eliminated the feature.) Because such crystal defects produce strong strain fields that are not completely accounted for by our NV Hamiltonian fitting procedure, their positions must be recorded for each diamond and the affected pixels excluded from measured magnetic field maps.

## 3.2 Illustration of Sensitivity and Spatial Resolution

To establish performance benchmarks for rock magnetometry with the QDM, we obtained maps of magnetic fields produced by remanent magnetization in a 30 µm thin section from the eucrite ALHA81001

(Figure 2d; full vector dataset in Supplementary Figure S6). Large-scale metal-silicate differentiation on the eucrite parent body, likely the asteroid Vesta, led to the generation of a core dynamo and remanent magnetization like that recorded by ALHA81001 [Fu *et. al.* 2012]. ALHA81001 and other eucrites exhibit very weak remanent magnetization due to the low abundance of metal grains [Rochette *et. al.* 2003], making ALHA81001 a stringent test for QDM sensitivity. At the same time, the QDM can aid in the identification of the main ferromagnetic carrier phases in ALHA81001. Based on probable observed Curie temperatures at 320 - 350°C and >700°C, Fe sulfide and Fe metal have been identified as the most likely remanence carriers [Fu *et. al.* 2012]. However, scanning electron microscopy only located Fe sulfides with the composition of troilite, which is paramagnetic at room temperature, and did not isolate grains of Fe metal with high certainty.

Similar to the experiments on Allende, we placed diamond D1 directly on the surface of the ALHA81001 thin section. (In this case, however, no IRM was imparted to the sample prior to measurement.) Magnetic fields due to remanent magnetization were imaged by VMM in a series of partially-overlapping FOVs, each with area 1.5 mm × 0.6 mm, with bias $B_0$ = 1.44 mT oriented along unit vector (0.17 $\hat{x}$ + 0.94 $\hat{y}$ + 0.30 $\hat{z}$) in the sensor coordinate system. We used a total averaging time of ~4 × $10^4$ s for each FOV, with an optical excitation intensity ~$10^6$ W/m$^2$ and imaging pixel size of (2.4 μm)$^2$. Spatial filters were applied in post-processing, including a 5 μm FWHM Gaussian low-pass to improve the signal-to-noise ratio (SNR), and a 200 μm Butterworth high-pass to eliminate offsets in each FOV associated with our nonzero residual bias $B_0^{(r)}$. After this procedure, we obtained a magnetic noise floor of ~20 nT RMS, estimated by calculating the pixel-wise magnetic field standard deviation for image regions where no sources were present. The sharpest magnetic features observed were on length scales of 5 – 10 μm and were not blurred significantly by filtering (Figure 2d, top center inset). To distinguish between features produced by sources within the sample and those due to strain features or magnetic contaminants on the diamond surface, we rotated the diamond 90° and repeated the measurement for each FOV (Supplementary Figure S7). Spurious strain-induced patterns were manually removed from the image, resulting in several blank areas in the map shown in the figure. These could be filled in by translating or rotating the diamond and measuring again, but were retained here for illustrative purposes. The measurement noise floor corresponds to an image-area-normalized magnetic field sensitivity of ~20 μT·μm/Hz$^{½}$ at long averaging times over a ~1 mm$^2$ FOV, which is typical VMM performance for the present QDM for diamonds D1 and D2 with NV layer thickness $t_{NV}$ ≈ 10 nm (Supplementary Figure S8).

Guided by the high-resolution QDM magnetic maps, we identified three strongly-magnetized locations in ALHA81001 (Figure 2d, top left inset) for compositional analysis using energy dispersive spectroscopy (EDS). These measurements were performed on a Zeiss EVO 60 environmental scanning electron microscope (ESEM) at the American Museum of Natural History (AMNH). Although the ≤1 μm diameters of the grains of interest precluded quantitative measurements of composition, we found that magnetization is spatially associated with (i) an Fe-bearing phase with no other transition metals or sulfur, interpreted to be Fe metal, and (ii) an Fe-Ni-Cr-bearing phase, interpreted to be non-stoichiometric chromite (Supplementary Figure S9). The identification of Fe metal is consistent with the >700°C Curie temperature observed during thermal demagnetization [Fu *et. al.* 2012]. Meanwhile, chromite in terrestrial rocks have been observed to demagnetize between 300°C and 400°C [Kumar *et. al.* 1984], again consistent with the demagnetization behavior of ALHA81001. If this interpretation is correct, ALHA81001 would be the second achondrite after the martian orthopyroxenite ALH 84001 where spatial correlation of magnetic field sources suggests chromite as a ferromagnetic carrier [Weiss *et. al.* 2002]. Because our EDS characterization of the Fe-Ni-Cr-bearing phase lacks the spatial resolution necessary to quantify the composition, we cannot make a positive identification of chromite or rule out the possibility of unresolved, nanoscale intergrowths of, for example, Fe metal that carry the observed magnetization. Future TEM-based work is required to address these ambiguities. Even so, a Fe-Ni-Cr-bearing phase likely crystallized during the primary cooling of ALHA81001 on its parent body as it is not consistent with terrestrial weathering products [Buchwald and Clark 1989]. As such, our QDM measurements provide further support for the extraterrestrial origin of magnetization on ALHA81001.

The QDM magnetic sensitivity demonstrated in our map of the ALHA81001 thin section can likely be improved for many applications by increasing the NV layer thickness, $t_{NV}$. The transverse spatial resolution of the QDM is generally limited by the sensor-sample standoff distance $d_{s-s}$, and the NV layer thickness can be increased up to a significant fraction of $d_{s-s}$ without adversely affecting the resolution (Supplementary Figure S10). Proposed QDM paleomagnetic imaging applications require spatial resolution of ~1 – 100 μm to resolve ferromagnetic grain populations in a wide range of geological samples, suggesting that the thin NV layers of D1 and D2 (with $t_{NV} \approx 10$ nm) should be replaced with thicker NV layers to enable more sensitive detection. For example, a diamond with NV layer thickness $t_{NV} = 10$ μm and equal NV density to a diamond with $t_{NV} \approx 10$ nm will produce ~$10^3$ times higher fluorescence at the same illumination intensity, yielding an SNR improvement of ~30 for photon shot-noise limited detection. Thick NV layer diamonds can be produced by doping with high N concentration during the last stage of CVD growth, followed by electron irradiation and annealing to create NV centers. To date, we have tested one such thick-NV-layer diamond, D3, and applied it successfully to large-FOV magnetic imaging studies of zircons (see discussion in section 3.4 below). We have yet to optimize the QDM for the high-throughput fluorescence collection needed to obtain maximum SNR improvement from thick-NV-layer diamonds when using smaller FOVs.

To achieve optimal spatial resolution and sensitivity in future QDM imaging applications, geological sample thickness, $t_{samp}$, will also be an important consideration. Because inversion of magnetic field maps to obtain volume distributions of dipolar sources in the sample is in general an ill-posed problem, the highest confidence determination of magnetization distribution uses samples with large area relative to their thickness [Lima et. al. 2013]. This requirement imposes a practical bound on QDM standoff distance and hence the imaging resolution, $d_{s-s} \gg t_{samp}$. The sample thickness constraint is relevant to the ALHA81001 field maps, where the QDM is most sensitive to sources in the top 5 μm. For the chosen QDM resolution, a much thinner rock section would be preferable for unambiguous determination of sample magnetization.

### 3.3 Accuracy of Magnetic Field Measurements

Absolute accuracy of magnetic field maps is necessary for correct determination of sample remanence and for proper comparison of magnetic measurements made in different devices. When ODMR spectra are acquired for both the $\Delta m_s = +1$ and $\Delta m_s = -1$ transitions, the accuracy of the QDM is relatively insensitive to temperature variations and other systematic effects that result in equal frequency shifts for the two resonances. Nevertheless, the linearity of magnetic field-induced spectral line shifts may break down at small $B_0$, when the Zeeman term in the NV Hamiltonian becomes comparable to nuclear hyperfine couplings and strain [Doherty et. al. 2013]. We therefore calibrated the QDM by incorporating an additional coil under the sample holder so as to produce uniform, well-defined fields perpendicular to the sensor surface (Figure 3a). Coil currents were driven using a diode laser driver and the coil geometry was precisely measured in an optical microscope, such that the expected magnetic field could be calculated with uncertainty < 1% (Supplementary Figure S11). We first obtained calibration curves of measured $B_z$ as a function of applied current for PMM (Figure 3b) and VMM (Figure 3c) using diamond D4 at both high and low bias fields ($B_0 = 18.6$ mT and $B_0 = 1.6$ mT, respectively). The measured slope agreed well with calculations (Figure 3d) to within the estimated uncertainty (grey band is 1 σ in figure), except in the case of VMM at low $B_0$. This deviation was likely due to strain inhomogeneity over the imaged region, which can cause line shifts as large as ~1 MHz in D4. The NV orientation with the smallest projection of $B_0$ may experience a Zeeman shift of only a few MHz in low-bias VMM, such that strain significantly broadens and/or shifts the spin transition frequencies for this orientation (Figure 3d, inset). This is an important potential drawback of VMM at low $B_0$.

Diamond strain and NV hyperfine effects may play an even greater role for CPMM imaging at low bias, $B_0$ < 0.2 mT. In this mode, we use only diamonds with NVs formed from $^{15}$N (nuclear spin $I$ = ½) implants, to avoid the spectral congestion and inhomogeneity associated with transitions involving $m_I$ = 0 states (for which the absence of hyperfine coupling makes strain the dominant interaction) in NVs formed from $^{14}$N ($I$ = 1). Nevertheless, sensor calibration is still essential to extract accurate magnetic field values, as demonstrated in our measurements on diamond D2 (Figure 3e). We observed a strong deviation from linearity in the QDM response for $B_z \approx 112$ µT, where the Zeeman and hyperfine energies are approximately equal, due to a transverse-strain induced anticrossing between energies of the $|m_s = \pm 1, m_I = \mp 1\rangle$ states. In addition, our spectral fitting algorithm was unable to determine independent line shape parameters for the transitions near degeneracy, resulting in a gap in our calibration. A second deviation from linearity for $B_z$ < 20 µT (Figure 3e, bottom inset) could also be attributed to strain. Because these nonlinearities are due to fixed material properties of the diamond chip, the calibration is repeatable (for a given FOV) and exact to within the accuracy of the applied calibration field.

## 3.4 Reconfigurable Multi-scale Imaging for Magnetic Surveys

The magnetic signal-to-noise ratio (SNR) of the QDM varies inversely with the laser excitation spot diameter in the shot-noise detection limit, for fixed laser power. It is therefore technically straightforward to alternate between rapid, wide-FOV magnetic mapping and targeted, sensitive acquisition in a small region of interest, simply by changing lenses in the optical excitation and fluorescence collection paths. This capability is advantageous in applications where a large sample area must be quickly screened to find magnetic features of interest. To demonstrate, we surveyed a group of detrital zircons from the Jack Hills region in Western Australia (Figure 4). The Jack Hills zircons are the oldest known samples of the Earth's crust, with ages ranging up to 4.38 billion years[Holden *et. al.* 2009]. As such, their magnetizations may potentially contain the earliest paleointensity records of the geodynamo [Tarduno *et. al.* 2015]. However, the presence of multiple metamorphic and metasomatic events in the locality suggests that their ferromagnetic phases may not be primary, but deposited in cracks and on the zircon exterior well after the formation of the zircon [Weiss *et. al.* 2015, 2016]. This possibility can be assessed by resolved magnetic imaging of zircons to determine the spatial distribution of their magnetization.

With this goal, we imaged the magnetization of 257 zircons using the QDM. The zircons were extracted from the host quartz-pebble conglomerate using a Frantz Model LB-1 Magnetic Separator at Australia National University (in which the grains were exposed to fields up to 1.6 T), washed in alcohol and then mounted in an epoxy disk and polished to approximately their midsections. They were then dated with Pb-Pb chronometry using secondary ion mass spectrometry at UCLA following the methods of Holden et al. (2009). Following this this, they were given an isothermal remanent magnetization (IRM) in a field of 140 mT, oriented perpendicular to the plane of the epoxy disk and then measured with VMM. (Also see the supplementary materials of [Fu *et. al.* 2017] for the discussion of a similar QDM imaging procedure applied to zircons from the Bishop Tuff.) Our analyzed samples included 22 grains with Pb-Pb ages greater than 3.9 Ga.

We first acquired coarse VMM magnetic maps of a subset (*n* = 257) of the zircons over four large FOVs (3.6 mm × 3.6 mm) with diamond D3, using an achromatic doublet lens (focal length *F* = 5 cm, numerical aperture $NA$ = 0.25) for fluorescence collection. The tube lens for all zircon-imaging experiments was also an achromatic doublet, $F$ = 15 cm. Each FOV was acquired using pixel size (8.7 μm)$^2$ and averaging time $1.8 \times 10^4$ s, yielding a noise floor of ~70 nT RMS. (Two of these FOVs are indicated in Figure 4a.) At this resolution, 147 out of 257 of zircons mapped produced magnetic signals distinguishable from background, although 122 of these were due to sources distributed on the exterior surface of the zircon. To assign magnetic features, co-registered magnetic and reflected-light images were overlaid and inspected visually (Figure 4 b,d). Magnetic sources were defined to be on the exterior of a zircon if the center of the dipole field pattern fell within ~20 μm of the boundary of that zircon in the reflected light image. The mix of magnetization directions observed from the zircon ferromagnetic sources is due to the combined effects of strong fields during magnetic separation and a weaker, 140 mT IRM in the out-of-plane direction.

We then changed to a higher magnification aspherized achromat lens ($F$ = 1.4 cm, $NA$ = 0.45) for light collection, and focused the excitation laser spot down to an area of 1.6 mm × 1.2 mm. This allowed us to zoom in on smaller FOVs that contained a high fraction of zircons dated to > 3.9 Ga. We acquired maps of six such FOVs using a pixel size of (3.6 μm)$^2$ and averaging time of $1.8 \times 10^4$ s, resulting in a noise floor of ~25 nT RMS. Field maps at the higher resolution were consistent with those of the original four large FOVs, but more magnetic features were visible due to the improved sensitivity and spatial resolution (Figure 4 c,e). Of the 78 zircons measured under these conditions, 71 now had detectable magnetic signatures, with 52 of those clearly attributable to sources on the exterior of the zircon. These results contrast with our QDM study of young (767 ka) relatively unaltered zircons from the Bishop Tuff [Fu *et. al.* 2017], which found that most of the magnetization is in the interior of the grains. Therefore, the present results emphasize that the natural remanent magnetization (NRM) in Jack Hills zircons [Tarduno *et. al.* 2015] could be far younger than their crystallization ages. In particular, because Tarduno 2015 measured the bulk NRM of zircons rather than imaging the NRM distribution, their data do not constrain where in the grains the magnetization carriers reside. A caveat is that these QDM data are images of IRM rather than NRM. In any case, these results also illustrate the flexibility of the QDM for imaging at different scales, from coarse surveys of large (~10 – 20 mm$^2$) FOVs, to detailed maps of smaller FOVs (~1 mm$^2$) with improved sensitivity, and hence better fidelity in detection of weak magnetic sources.

## Conclusion

We constructed a new instrument, the quantum diamond microscope (QDM), for imaging magnetic fields from room-temperature geological samples with spatial resolution ~5 μm. The QDM also provides optical images of the sample that are spatially-correlated with the magnetic images. The device can be operated in three modes, including (i) a vector (three-axis) magnetic imaging mode, (ii) a projective (single-axis) mode optimized to improve magnetic field sensitivity by a factor of 2 – 3, and (iii) a single-axis mode using circularly-polarized microwaves that allows operation at low bias $B_0$ < 10 μT. The first two modes (VMM and PMM) provide superior absolute accuracy in magnetic field estimation (~1% without calibration), particularly at $B_0$ > 10 mT, but rely on precise bias reversal (with $B_0^{(r)}$ < 0.1 μT) to distinguish between paramagnetic and ferromagnetic sources. The third mode (CPMM) avoids the bias-reversal constraint by operating at low ambient field, but requires careful calibration to account for strain-induced shifts in the NV ODMR spectra. The typical image-area-normalized magnetic field sensitivity of the present QDM is ~20 μT·μm/Hz$^{½}$ for a 1 mm × 1 mm FOV, and scales linearly with the diameter of the FOV for fixed laser power. The best demonstrated noise floor is ~20 nT RMS. We recently began to work with diamond chips that possess considerably thicker NV layers than were previously available (~10 μm rather than ~10 nm), which are expected to provide a sensitivity improvement of ~30× with the implementation of a faster camera and data acquisition system. We used the QDM to image magnetic fields from a variety of magnetically heterogeneous rock samples, and confirmed that we can distinguish populations of ferromagnetic carriers separated by <10 μm. In particular, QDM imaging shows that magnetization carriers in ancient zircon crystals from the Jack Hills of Western Australia are largely confined to the exteriors of most grains, suggesting that their ferromagnetic minerals are secondary in origin and therefore that they do not retain pristine records of the earliest history of the dynamo.

A number of technical improvements are planned for future QDM systems, including (i) tools for rapid sample exchange and alignment, to increase measurement throughput and facilitate QDM imaging interspersed with multiple rounds of AF or thermal demagnetization, (ii) improved heatsinking and thermal stabilization, enabling the use of higher laser power for better QDM sensitivity, and (iii) development of micron-scale magnetic standard samples, to allow quantitative comparison between QDM magnetic maps and those obtained in other instruments at different standoff distances.

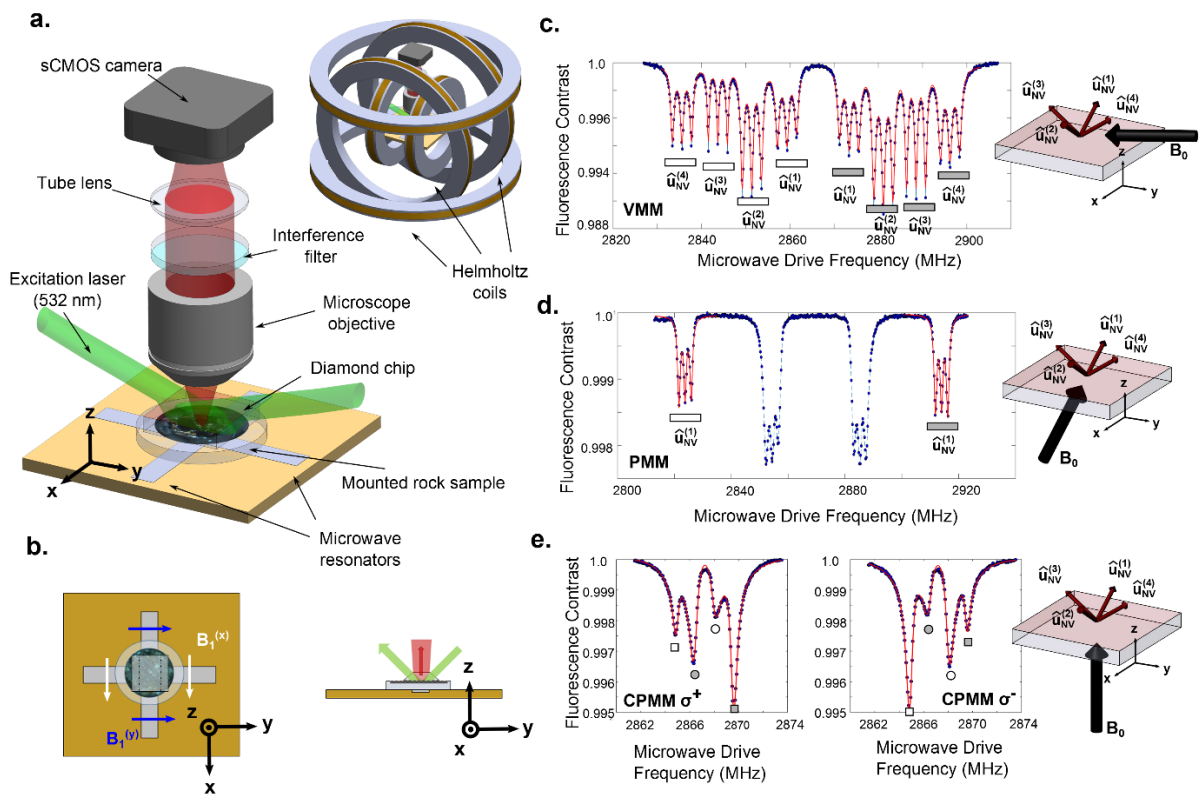

**Figure 1: Rock magnetic imaging using a quantum diamond microscope (QDM). (a)** Schematic of QDM configuration. The rock sample is placed on a glass slide near the focus of a microscope objective. A diamond chip is positioned above it, with a dense layer of nitrogen-vacancy (NV) centers facing down. The NV centers are interrogated using green laser light and continuous microwave driving, with the microwaves delivered by strip-line resonators below the sample mount. Red fluorescence from the NV centers is imaged onto a scientific CMOS camera to map the sample magnetic fields. Inset shows full QDM including three orthogonal pairs of Helmholtz coils that control the bias field $B_0$ at the NV center layer. **(b)**

Isolated top and side views of the sample mount plus microwave strip-lines, which are driven to produce a resonant field ($B_1$) to manipulate the NV spins. When the strip-lines are driven in phase, $B_1$ is linearly polarized along $\hat{x} + \hat{y}$. When the strip-lines are driven out of phase, $B_1$ is circularly polarized about the $z$ axis (with polarization vector $\hat{x} \pm i\hat{y}$). Laser light impinges on the NV center layer at the bottom of the diamond chip, and NV fluorescence is imaged through the top of the diamond. **(c)** Characteristic NV optically-detected magnetic resonance (ODMR) spectrum for vector magnetic microscopy (VMM). Inset shows the orientation of $B_0$, which projects unequally onto the [111] diamond lattice directions so that transitions for all four NV orientations are resolved. Each orientation allows two spin transitions, $\Delta m_s = \pm 1$, which are further split into triplets due to hyperfine coupling to the 14N nucleus of the NV center. The $\Delta m_s = +1$ (or $\Delta m_s = -1$) triplets are indicated with grey (or white) bars. During data acquisition, the $B_1$ frequency is swept across all resonances, the NV fluorescence is measured at each frequency (blue dots) and a portion of the spectrum is fit (red curve) to determine the local vector magnetic field. **(d)** ODMR spectrum for projective magnetic microscopy (PMM). $B_0$ is parallel to one of the [111] directions, distinguishing it spectrally, while the other three NV orientations are degenerate. The $B_1$ frequency is scanned and line shapes are fit (red curves) only for the aligned transitions, enabling rapid acquisition of single-axis magnetic field data. In a real experiment, data points corresponding to the non-aligned transitions would not be acquired – they are included here for illustrative purposes only. **(e)** ODMR spectrum for circularly-polarized magnetic microscopy (CPMM). The NV centers are formed from $^{15}$N, resulting in hyperfine doublets instead of the triplets associated with $^{14}$N. $B_0$ is applied perpendicular to the chip surface, with equal projection on all four [111] directions, making the four pairs of NV transitions degenerate. Also, $B_0$ is small, such that the electronic Zeeman shift is weaker than the nuclear hyperfine coupling. Transitions for which the nuclear spin state is $m_I = +\frac{1}{2}$ (or $m_I = -\frac{1}{2}$) are indicated with squares (or circles). The $\Delta m_s = +1$ (or $\Delta m_s = -1$) transitions are indicated with grey (or white) markers. In the first (second) panel, $B_1$ has right- (left-) circular polarization about the $B_0$ axis, favoring excitation of the $\Delta m_s = +1$ ($\Delta m_s = -1$) transitions. Data acquisition consists of alternating between circular-polarization orientations while scanning the $B$ frequency and fitting line shapes over the full spectrum (red curve).

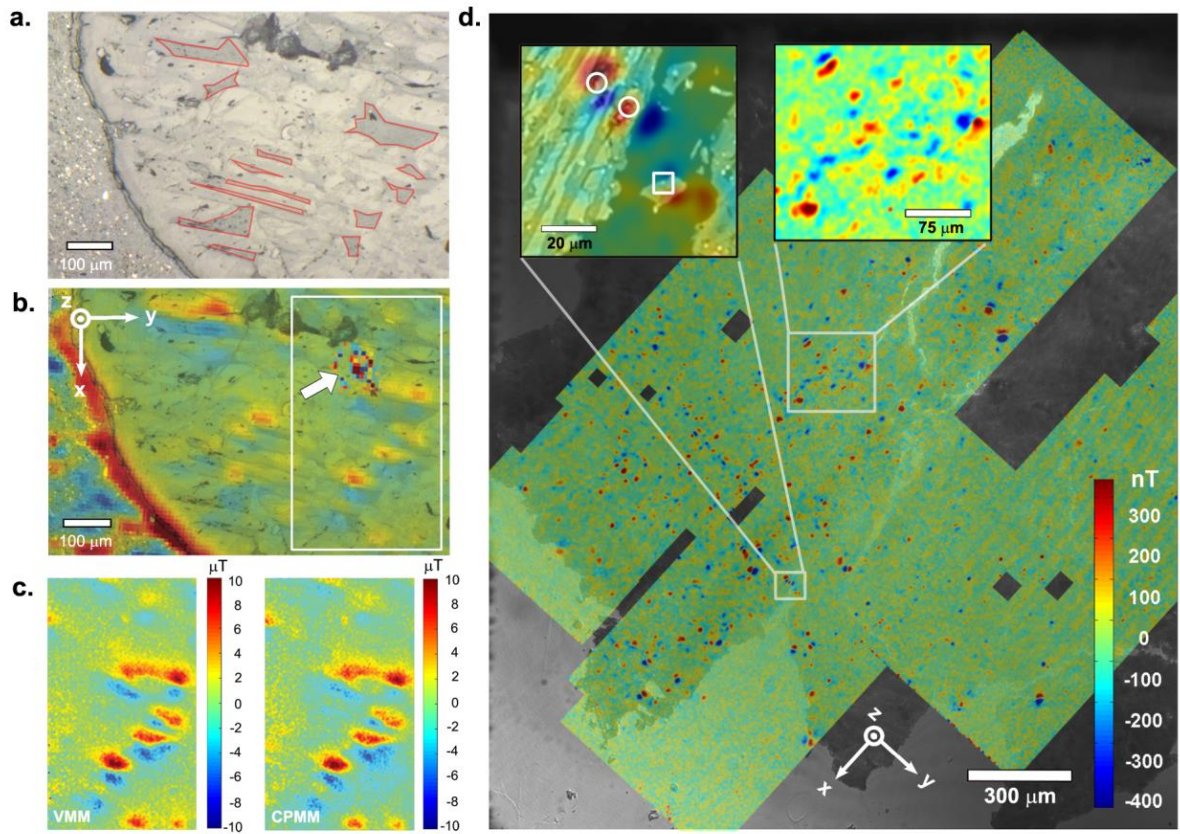

**Figure 2: QDM magnetic field maps of remanent magnetization and spatially-correlated optical images of meteorites. (a)** Reflected-light image (plane-polarized illumination) of a 30 µm thin section of a chondrule from the Allende CV3 chondrite. Mesostasis regions are outlined in red. **(b)** Magnetic map of the $B_z$ (out of plane) component of the field distribution produced by the sample, acquired in VMM and overlaid on the same reflected-light image. White arrow indicates a distortion of the field map due to a localized diamond defect. Such defects are fixed in the diamond volume and may be removed from the FOV by translating the sensor relative to the sample [as in (c)]. **(c)** Comparison of VMM (left panel) and CPMM (right panel) field maps of the sub-region of (b) enclosed in the white box. The residual bias after field reversal ($B_0^{(r)}$) has been subtracted from the VMM map; the full bias ($B_0$) has been subtracted from the CPMM map. The resulting magnetic field images are qualitatively similar, but peak recorded field intensities differ by up to ~10%, illustrating the need for independent calibration of CPMM. **(d)** Overlay of a $B_z$ (out-of-the-plane) VMM magnetic field map and reflected light photomicrograph of a 30 µm thin section from the weakly-magnetized eucrite ALHA81001. The magnetic map is composed of four tiled and partially overlapping FOVs, with several small regions removed due to known strain defects in the diamond. The QDM data have spatial resolution ~5 µm and a noise floor of ~20 nT RMS. Inset Top Left: Backscattered electron (BSE) image of a small region of the sample, overlaid with the QDM-acquired $B_z$ map of the same region. The magnetic map was used to guide energy dispersive spectroscopy (EDS) analysis at select locations with strong magnetic sources, which found elemental abundances associated with chromite (white circles) and metallic iron (white square). Inset Top Center: Zoomed-in region of the QDM magnetic field image (with limits indicated by solid white box), demonstrating the spatial resolution of the instrument.

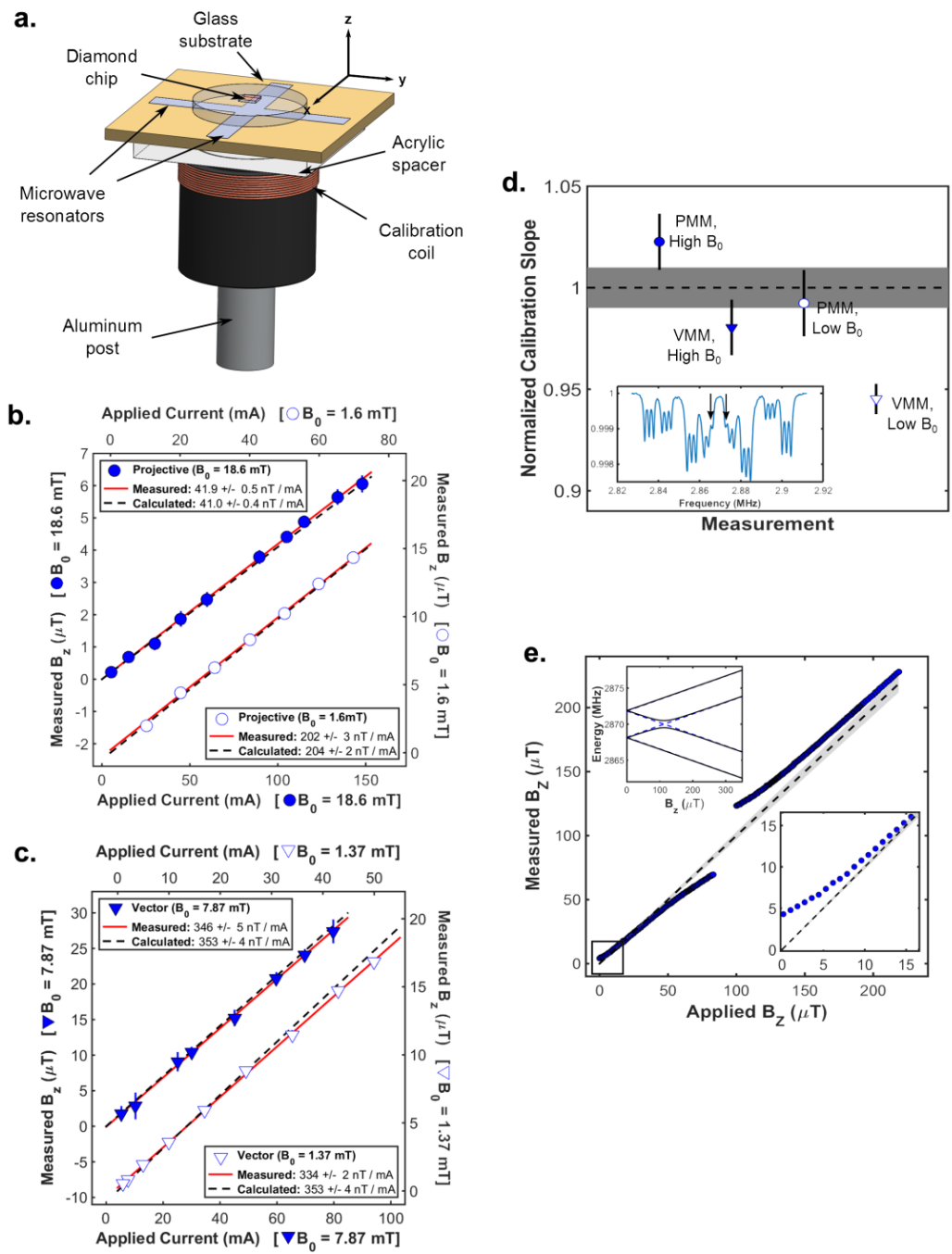

**Figure 3: Absolute magnetic field calibration of the QDM. (a)** QDM calibration setup schematic. A precisely characterized coil, placed below the diamond chip, generated known magnetic fields at the NV sensing layer. The ODMR signal was averaged over a 1.0 mm × 0.5 mm FOV, and the resulting measured magnetic fields compared against the calculated coil field. **(b)** Magnetic field values obtained in PMM as a function of coil current, for both high bias ($B_0$ = 18.6 mT, closed circles) and low bias ($B_0$ = 1.6 mT, open circles). Linear fits to the data (red lines) are consistent with the calculated fields (black dashed lines) for each current value. **(c)** Magnetic field values obtained in VMM as a function of coil current, for both high

bias ($B_0$ = 7.87 mT, closed triangles) and low bias ($B_0$ = 1.37 mT, open triangles). A linear fit to the data (red lines) is consistent with the calculated fields (black dashed lines) for the high bias case, but differs by more than the estimated uncertainty in the calculation for the low bias case. **(d)** Summary of fit slopes from (b) and (c), normalized to the calculated value. Grey band indicates uncertainty (1 $\sigma$) in the calculated slope. Error bars on the individual measured slopes are also 1 $\sigma$. Inset: Characteristic low-$B_0$ VMM spectrum, with the innermost peaks (black arrows) broadened by transverse strain. **(e)** CPMM calibration, showing measured magnetic field vs. expected field as calculated from the applied current. Black dashed line is the ideal calibration with unit slope; grey band indicates uncertainty in the calculation. Measurements (blue data points) differ significantly from the ideal curve due to a transverse strain-induced avoided crossing. Inset Bottom Right: Zoomed-in data for $B_z$ < 10 µT. The measured magnetic field is nonzero at zero applied field due to a finite splitting induced by transverse strain. Inset Top Left: Calculated energy structure of the $m_s \neq 0$ NV spin states as a function of applied magnetic field for a nominal transverse strain parameter of 0.5 MHz (black curve), compared to the zero-strain case (dashed blue curve). Strain results in an avoided crossing, as well as a small additional splitting of the energies at zero magnetic field.

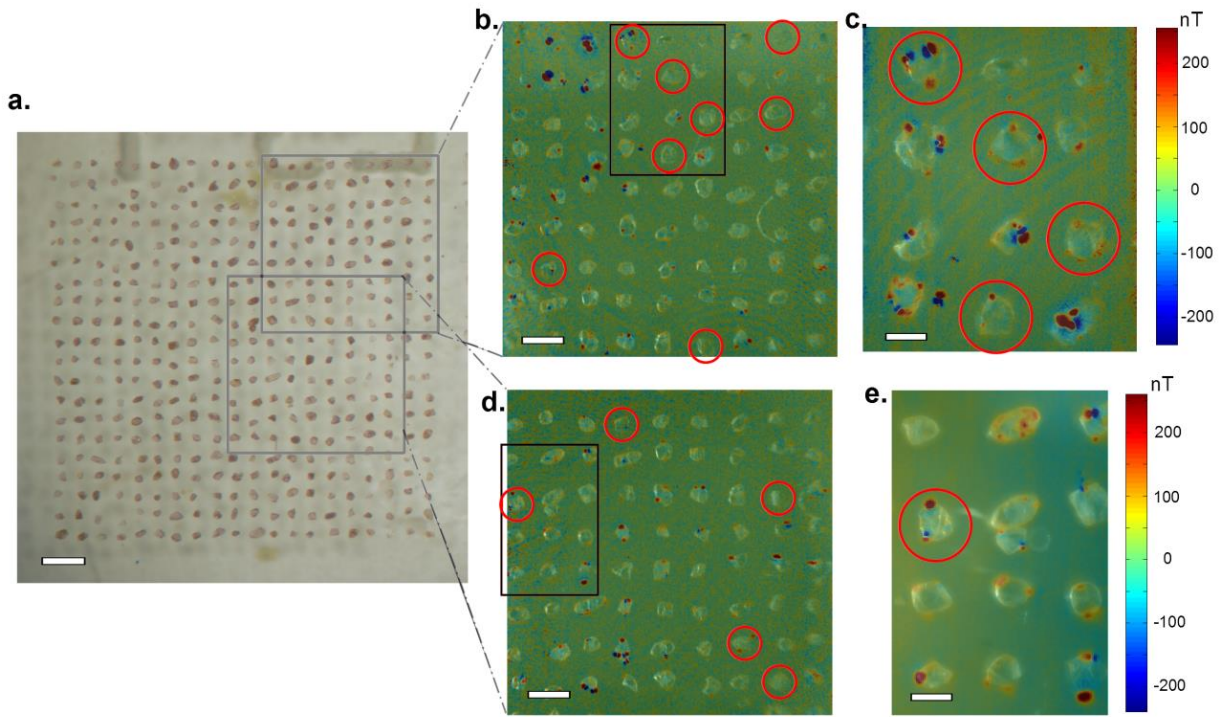

**Figure 4: Multi-scale QDM survey of magnetic sources in Jack Hills zircons. (a)** Reflected-light image of a matrix of zircons from the Jack Hills of Western Australia, embedded in epoxy. Grey boxes show two characteristic FOVs for which VMM magnetic field maps were obtained. Scale bar is 1 mm. **(b)** Reflected light image of upper FOV, with the $B_z$ (out-of-plane) component of the VMM magnetic map overlaid. The magnetic data acquisition was optimized to allow rapid imaging of a large (~13 mm²) field of view. Red circles indicate zircons > 3.9 Ga in age. Scale bar is 500 µm. Color scale (not shown) is -500 nT - 500 nT. **(c)** Reflected light image of zoomed-in region in upper FOV, with the $B_z$ (out-of-plane) component of the VMM magnetic map overlaid. The magnetic data acquisition was optimized for magnetic field sensitivity and spatial resolution. Magnetic features qualitatively agree with those observed in (b), although more features are visible due to the improved sensitivity. Scale bar is 200 μm. Color scale is -250 nT - 250 nT. **(d)** Reflected light image of lower FOV, with the $B_z$ (out-of-plane) component of the VMM magnetic map overlaid. Red circles indicate zircons > 3.9 Ga in age. Scale bar is 500 μm. Color scale (not shown) is -500 nT - 500 nT. **(e)** Reflected light image of zoomed-in region of lower FOV, with the $B_z$ (out-of plane) component of the VMM magnetic map overlaid. Scale bar is 200 μm. Color scale is -250 nT - 250 nT.

## Acknowledgments, Samples, and Data

We thank the NASA Emerging Worlds and NASA Planetary Major Equipment programs (grant # NNX15AH72G), the NSF Integrated Support Promoting Interdisciplinary Research and Education (INSPIRE) program (grant #EAR 1647504), the NSF Electronics, Photonics and Magnetic Devices (EPMD) program (grant # 1408075), and the DARPA Quantum Assisted Sensing And Readout (QuASAR) program (contract # HR0011-11-C-0073) for support. PK acknowledges support from the IC Postdoctoral Research Fellowship Program. B.P.W, E.A.L., and R.R.F. thank Thomas F. Peterson Jr for a generous gift.  We also thank T. M. Harrison for providing the Jack Hills zircons used for QDM imaging.

# Supplementary Information

## Supplementary Table S1: Diamond Chip Catalog

| Diamond | Chip Size | NV Layer Type | NV Layer Depth | NV Layer Thickness | NV Density |
|---|---|---|---|---|---|
| D1 | $2 \times 2 \times 0.5$ mm$^3$ | $^{14}$N implant at 14 keV | 20 nm | 10 nm | $\sim 2 \times 10^{11}$ cm$^{-2}$ |
| D2 | $2 \times 2 \times 0.5$ mm$^3$ | $^{15}$N implant at 14 keV | 20 nm | 10 nm | $\sim 8 \times 10^{10}$ cm$^{-2}$ |
| D3 | $4 \times 4 \times 0.5$ mm$^3$ | $^{14}$N CVD doped layer | 6.5 µm | 13 µm | $\sim 2 \times 10^{17}$ cm$^{-3}$ |
| D4 | $4 \times 4 \times 0.5$ mm$^3$ | $^{14}$N CVD doped layer | 2 µm | 4 µm | $\sim 2 \times 10^{17}$ cm$^{-3}$ |

**Table S1:** Comparison of dimensions and NV properties for the four diamond chips used in this study. NV layer thickness and mean NV layer depth for ion-implanted samples D1 and D2 were estimated from numerical calculations using the Stopping Range of Ions in Matter (SRIM) software package, for an ion implantation energy of 14 keV. NV layer thickness and mean NV layer depth for the grown-layer samples D3 and D4 are quoted according to supplier (Element Six) specifications. For all diamonds, NV densities were estimated using relative fluorescence intensity measurements in a home-built confocal microscope, and are believed to be correct to within a factor of $\sim$2. Densities are quoted in units of surface density (cm$^{-2}$) for D1 and D2, and units of volume density (cm$^{-3}$) for D3 and D4.

## Supplementary Figure S2: Heating Due to Laser Illumination:

Sample heating by the green laser used to excite NV fluorescence could potentially complicate QDM paleomagnetic measurements by causing thermal remagnetization. Both the diamond and the substrate absorb laser light, with the resulting temperature increase depending on how efficiently the heat generated from this absorption is dissipated. To assess the severity of this heating problem, we measured the diamond temperature via ODMR [Gruber *et. al.* 1997] while varying the green laser power, with the diamond placed on one of several characteristic substrates. We assume that the diamond and substrate temperatures are approximately equal, due to their physical contact and the high thermal conductivity ($2 - 3 \times 10^3$ W/m·K) of diamond.

As the diamond temperature changes, all of the NV resonance frequencies shift uniformly by -74.2 kHz/K [Acosta *et. al.* 2010]. We used Vector Magnetic Microscopy (VMM) for this assessment, though Projective- (PMM) and Circular-Polarized Magnetic Microscopy (CPMM) also work. To avoid excessive heating, one can decrease the laser power (at the expense of increased averaging time to obtain the same magnetic field sensitivity), mount the diamond sensor on a heatsink within the QDM, or separate the sample and diamond with an air gap. Because the diamond temperature is recorded in the ODMR spectrum, it is a straightforward precaution to ramp the laser power gradually up from zero at the start of a measurement, until the maximum safe temperature for a given sample is reached

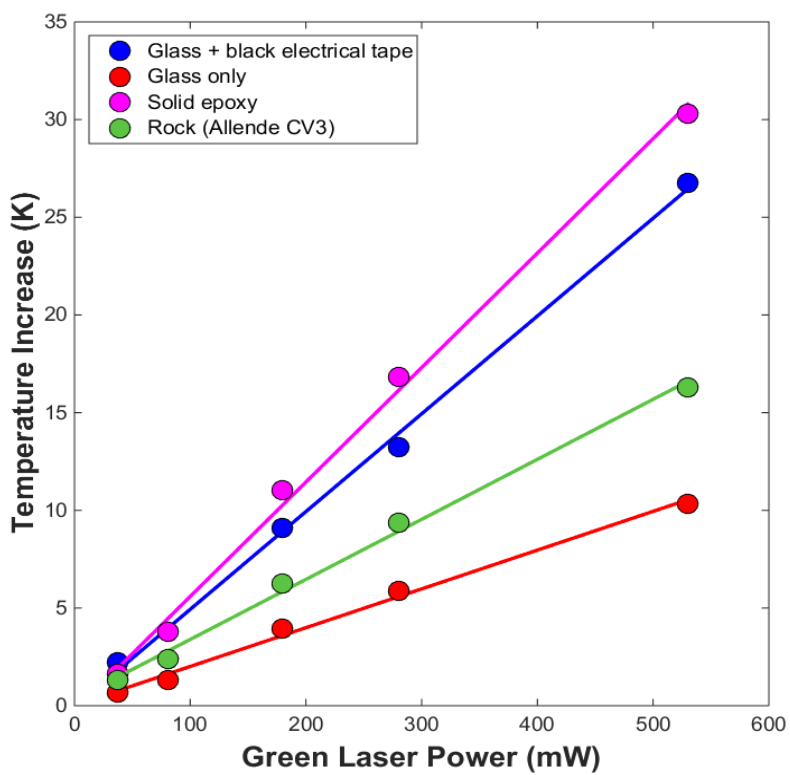

**Figure S2:** Measured diamond temperature increase relative to room temperature, as a function of green laser power, for various substrates. Circles indicate measured data points, with uncertainties approximately equal to the circle size; solid lines are linear fits to the data. Data were acquired using diamond D4 with a 1.0 mm × 0.5 mm FOV. The epoxy substrate was the same as that used to mount the zircons in Fig. 4 of the main text. The rock substrate was the same section of the Allende CV3 chondrite shown in Fig. 2a-c of the main text.

## Supplementary Figure S3: Introduction to NV Centers in Diamond

The nitrogen-vacancy (NV) center consists of a substitutional N atom adjacent to a vacancy in the diamond lattice. It is a color center, with localized electronic and spin states that can be probed by optical and microwave spectroscopy. The NV center has attracted much recent interest for quantum information and sensing applications, and is described extensively in a number of recent reviews [Doherty *et. al.* 2013, Schirhagl *et. al.* 2014, Rondin *et. al.* 2014].

The electronic ground state of the NV center is a spin triplet. In the absence of an applied magnetic field, spin-spin interactions in the ground state electronic wave function cause the states with one unit of spin angular momentum projected on the N – V axis (denoted $m_s$ = ±1) to have higher energy than the state with zero spin projection along the N – V axis ($m_s$ = 0). This energy difference, called the zero-field splitting (ZFS), is approximately $f_{ZFS} \approx$ 2.87 GHz in frequency units (Figure S3a). When a magnetic field is applied with finite projection along the N – V axis, the Zeeman interaction lifts the degeneracy of $m_s$ = ±1, causing the $m_s$ = +1 state to move to higher energy, and the $m_s$ = -1 to move to lower energy. A microwave field with frequency $f$ swept in a narrow range around 2.87 GHz resonantly excites two spin-flip transitions ($m_s$ = 0 → $m_s$ = -1 at $f < f_{ZFS}$, and $m_s$ = 0 → $m_s$ = +1 at $f > f_{ZFS}$). The frequency difference between the two resonances is linearly proportional to the applied magnetic field, with proportionality constant (2 $g$ $\mu_B$), where g = 2 is the electron Landé factor and $\mu_B \approx$ 14 GHz/T is the Bohr magneton.

This simple picture of NV spin transitions is complicated slightly by hyperfine interactions between the NV electronic spin and the N nuclear spin. The nitrogen from which the NV center is formed can be one of two isotopes: $^{14}$N, with nuclear spin $I$ = 1, or $^{15}$N, with nuclear spin $I$ = ½. Because the natural abundance of $^{14}$N is ~0.996, we generally consider only $^{14}$N, unless the diamond chip is grown or implanted with an isotopically-enriched source of $^{15}$N. For $^{14}$N, spin-spin interactions between the nucleus and the electrons cause the $m_s$ = ±1 states each to split into 3 energy levels, corresponding to the projection of the nuclear spin along the N – V axis, $m_I$ = -1, $m_I$ = 0, or $m_I$ = 1. This splitting is not present for the $m_s$ = 0 state, with the result that six energetically-distinct electronic spin transitions can now be driven: ($m_s$ = 0, $m_I$ = -1) → ($m_s$ = -1, $m_I$ = -1); ($m_s$ = 0, $m_I$ = 0) → ($m_s$ = -1, $m_I$ = 0) and so on. Note that the nuclear state $m_I$ is conserved by this magnetic dipole transition. The resulting microwave absorption spectrum for a single NV center consists of two hyperfine triplets, one due to $m_s$ = 0 → $m_s$ = -1 plus associated hyperfine interactions, the other due to $m_s$ = 0 → $m_s$ = +1 plus hyperfine. The situation is similar for NV centers formed from $^{15}$N, except: (i) only two $m_I$ states are available ($m_I$ = ±½), resulting in doublets instead of triplets in the absorption spectrum, and (ii) the magnitude of the observed splitting is somewhat different (3.03 MHz for $^{15}$N, compared to 2.16 MHz for $^{14}$N) due to different hyperfine interaction strengths.

At room temperature, green (532 nm) light can be used to drive optical transitions to vibronic sidebands of an excited triplet electronic state, which then preferentially decays by spontaneous emission of red light (638-800 nm) back to the ground electronic state. However, an alternate decay path exists, involving non-radiative relaxation via an electronic singlet manifold. This path is favored for the $m_s$ = ±1 magnetic sublevels compared to $m_s$ = 0, and results in net population transfer to the $m_s$ = 0 state of the triplet under continuous green illumination. Furthermore, if the NV is exposed to continuous green laser illumination and a microwave drive field with frequency swept near 2.87 GHz, competition between the radiative and non-radiative decay paths produces a decrease in fluorescence whenever the microwaves come into resonance with the ground state spin transition. This technique, called optically-detected magnetic resonance (ODMR), provides the basis for magnetic field sensing with the QDM.

The QDM diamond chips used in this study were cut such that the normal vector to the bottom (sensing) surface was parallel to the [100] crystallographic direction, and normals to the lateral faces of the chip were parallel to [110]. We define a coordinate system for all QDM magnetic images with x and y in the plane of the sensor, and z perpendicular to the plane (Figure S3b). In these coordinates, unit vectors for the [111] crystal directions, and therefore the allowed NV orientations, are as follows:

$$\hat{u}_{NV}^{(1)} = \left\{-\sqrt{\tfrac{2}{3}}, 0, \sqrt{\tfrac{1}{3}}\right\}, \quad \hat{u}_{NV}^{(2)} = \left\{\sqrt{\tfrac{2}{3}}, 0, \sqrt{\tfrac{1}{3}}\right\}, \quad \hat{u}_{NV}^{(3)} = \left\{0, \sqrt{\tfrac{2}{3}}, \sqrt{\tfrac{1}{3}}\right\}, \quad \hat{u}_{NV}^{(4)} = \left\{0, -\sqrt{\tfrac{2}{3}}, \sqrt{\tfrac{1}{3}}\right\}.$$

Here, we have chosen the z-projections of all NV orientation vectors to have the same sign; this is equivalent to a consistent definition of the $m_s = \pm 1$ states for all NV orientations under microwave drive fields that are circularly polarized with respect to z, and is relevant to our discussion of CPMM imaging (Supplementary Figure S5). The presence of four orientations of NV centers in the diamond chip results in four distinct pairs of triplets (for $^{14}$NV) or doublets (for $^{15}$NV) in the observed ODMR spectrum, assuming sufficiently different magnetic field projections along each NV orientation axis.

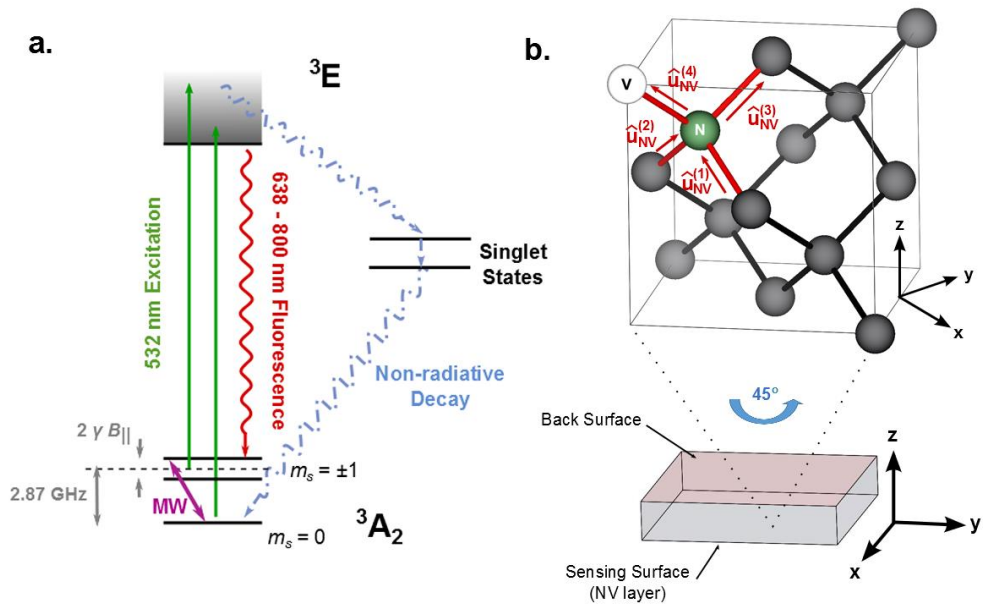

**Figure S3:** Magnetic imaging using NV-ODMR. **(a)** Level structure of the nitrogen vacancy center. Excitation of the NV electronic transition with 532 nm light while simultaneously driving the ground-state spin transitions leads to a decrease in fluorescence, as the $m_s = \pm 1$ states have a greater probability to decay via a non-radiative path than the $m_s = 0$ state. The reduced fluorescence can be used to detect the Zeeman-shifted spin resonances when the microwave frequency is swept, enabling magnetic field measurements. **(b)** Diamond coordinate system for QDM magnetic imaging. Geological samples are placed below the chip, immediately adjacent to the NV sensing layer, in the x-y plane. Expanded view shows a unit cell of the diamond lattice containing a single NV center. The cell is rotated 45° about z for improved visibility of component atoms. The NV center depicted lies along $\hat{u}_{NV}^{(4)}$, one of the four allowed orientations. The NV is most sensitive to magnetic fields parallel to its orientation vector.

## Supplementary Figure S4: ODMR Data Fitting Summary

To derive magnetic field maps from ODMR data, we follow a two-step procedure: (i) fit the ODMR spectrum for each pixel to extract the NV spin resonance frequencies, then (ii) use the resonance frequencies to determine the local vector (three-axis) or projective (single-axis) magnetic field at that pixel. The form of the fits differs somewhat for each mode of QDM operation.

After any binning and/or filtering, the processed ODMR data consists of a stack of $Q$ fluorescence images, each acquired at a different microwave drive frequency, and all $M \times N$ pixels in size. (We typically use $Q \sim 500 - 1000$, for VMM, or $Q \sim 100 - 200$ for PMM and CPMM, and $M, N \sim 100 - 300$ depending on the desired field of view.) We label the pixels with indices ($m = 1...M, n = 1...N, q = 1...Q$), such that the pixel with indices ($m,n,q$) has spatial coordinates $x = m \Delta x$, $y = n \Delta y$, in the fluorescence image acquired at microwave frequency $f_q$. For each pixel, we extract the $Q$-element array of fluorescence measurements at that point with varying frequency, and carry out a nonlinear least squares (Levenberg-Marquardt) minimization to determine the spectral parameters. Specifically, in the $(m,n)^{th}$ pixel, we minimize the sum of squares $\sum_{q=1}^{Q} \left( S_q - S(f_q, \vec{\alpha}_{m,n}) \right)^2$, where $S_q$ are the fluorescence data points, $f_q$ are the microwave frequencies at which they are acquired, and $S(f, \vec{\alpha}_{m,n})$ is the spectral fit function with a vector of fit parameters $\vec{\alpha}_{m,n}$ for the $(m,n)^{th}$ pixel. The spectral fit function $S$ is as follows:

For VMM:

$$S\left[f, \left(A_{1...24}, f_{1...8}^{(Res)}, \Gamma_{1...8}, C\right)_{m,n}\right]$$

$$= \sum_{j=1}^{8} \left[ \frac{A_{3j-2}}{\left(f - f_j^{(Res)} + d^{HF}\right)^2 + \Gamma_j^2} + \frac{A_{3j-1}}{\left(f - f_j^{(Res)}\right)^2 + \Gamma_j^2} \right.$$

$$\left. + \frac{A_{3j}}{\left(f - f_j^{(Res)} - d^{HF}\right)^2 + \Gamma_j^2} \right] + C$$

For PMM:

$$S\left[f, \left(A_{1...6}, f_{1..2}^{(Res)}, \Gamma_{1...2}, C\right)_{m,n}\right]$$

$$= \sum_{j=1}^{2} \left[ \frac{A_{3j-2}}{\left(f - f_j^{(Res)} + d^{HF}\right)^2 + \Gamma_j^2} + \frac{A_{3j-1}}{\left(f - f_j^{(Res)}\right)^2 + \Gamma_j^2} \right.$$

$$\left. + \frac{A_{3j}}{\left(f - f_j^{(Res)} - d^{HF}\right)^2 + \Gamma_j^2} \right] + C$$

For CPMM:

$$S\left[f, \left(A_{1..4}, f_{1..2}^{(Res)}, \Gamma_{1..2}, C\right)_{m,n}\right] =$$

$$\sum_{j=1}^{2}\left[\frac{A_{2j-1}}{\left(f - f_j^{(Res)} + d^{HF}/2\right)^2 + \Gamma_j^2} + \frac{A_{2j}}{\left(f - f_j^{(Res)} - d^{HF}/2\right)^2 + \Gamma_j^2}\right] + C$$

In the above fit functions, $d^{HF}$ is the hyperfine splitting of the NV, which is $d^{HF}$ = 2.16 MHz for $^{14}$NV (in VMM and PMM), and $d^{HF}$ = 3.03 MHz for $^{15}$NV (in CPMM) [Felton *et. al.* 2009]. The $A$, $f^{(Res)}$, $\Gamma$, and $C$ parameters describe (respectively) the amplitudes, resonance frequencies, linewidths and background offset of the Lorentzian lineshapes (c.f. Fig. 1c of the main text). Peak amplitudes $A$ are fit independently for each peak in a given hyperfine triplet (for VMM or PMM, using $^{14}$NV diamonds) or doublet (for CPMM, using $^{15}$NV diamonds) to account for varying degrees of NV nuclear spin polarization under continuous optical excitation [*Fischer* 2013]. One resonance frequency, $f^{(Res)}$, and one linewidth, $\Gamma$, are fit per triplet or doublet, and the fluorescence offset, $C$, is a global parameter for the full spectrum. Having extracted the spin resonance frequencies for NV centers in the $(m,n)^{th}$ pixel, we can now determine the magnetic field acting on them.

For single-axis measurements (PMM and CPMM), the projected magnetic field parallel to the NV orientation direction is straightforwardly estimated as $|B_\parallel| = \frac{|f_2^{(Res)} - f_1^{(Res)}|}{2\,g\,\mu_B}$, with $g$ = 2.003 [Felton *et. al.* 2009], and the Bohr magneton $\mu_B$ = 13.996 GHz/T. The sign of the bias field must be known to determine the sign of $B_\parallel$. We note that, in PMM, magnetic fields $B_\perp$ transverse to the NV axis result in frequency shifts scaling as $\Delta f \sim \frac{(\mu_B B_\perp)^2}{D_{ZFS}}$, with $D_{ZFS} \approx 2.87$ GHz. For typical geological samples that produce fields $B_\parallel, B_\perp \lesssim 10$ µT at the NV centers, the relative accuracy error in PMM due to the transverse contribution is $\lesssim 10^{-4}$.

For vector (three-axis) measurements (VMM), an additional nonlinear fit is required to extract the full magnetic field. The sum of squares to be minimized is now $\sum_{j=1}^{8}\left(f_j^{(Res)} - f^{(Res)}\left[\left(\vec{B}, D_{ZFS}^{(k)}\right)_{m,n}\right]\right)^2$. The model for the resonance frequencies, $f^{(Res)}$ depends on seven fit parameters: the Cartesian (three-component) magnetic field vector $\vec{B}$, and a diamond zero field splitting vector $D_{ZFS}^{(k)}$ with components $k$ = 1...4, each describing the temperature-dependent axial strain along one NV orientation. To evaluate $f^{(Res)}$, we separately diagonalize the NV spin Hamiltonian for each NV orientation $\hat{u}_{NV}^{(k)}$, $k$ = 1..4 (as defined in Supplementary Figure S3).

The Hamiltonian for orientation ($k$) is defined as:

$$H^{(k)} = D_{ZFS}^{(k)}\left(S_3^{(k)}\right)^2 + g\,\mu_B\,\vec{S}^{(k)} \cdot \vec{B}\ .$$

Here, $S_1^{(k)}$, $S_2^{(k)}$, and $S_3^{(k)}$ are the quantum spin operators for NV orientation ($k$), written in the usual way as 3 × 3 matrices for $S$=1. Also, $D_{ZFS}^{(k)}$ is the zero-field splitting (ZFS) energy for NV orientation (k), $g$ = 2.003, and $\mu_B$ is the Bohr magneton. For each orientation ($k$), the axial spin operator $S_3^{(k)}$ lies along $\hat{u}_{NV}^{(k)}$, with transverse spin operators $S_1^{(k)}$ and $S_2^{(k)}$ chosen arbitrarily to form a right-handed system. Transverse strain is not included in the model Hamiltonian, because (i) it is expected to be a small effect, since the transverse strain energy (~100 kHz – 1 MHz in our diamonds) is suppressed by $D_{ZFS}$ ($\approx$ 2.87 GHz) at low magnetic field (i.e. when $\mu_B \vec{B} \ll D_{ZFS}^{(k)}$, which is always true for our experiments); and (ii) the additional parameters required would cause the model to become under-constrained, necessitating acquisition of an additional magnetic field image at a known magnetic offset to solve for all parameters.

After diagonalizing the Hamiltonian $H^{(k)}$, the eigenvalues are sorted in ascending order as $E_0^{(k)}$, $E_1^{(k)}$, and $E_2^{(k)}$. This gives two transition frequencies, $f_{\Delta m_s=-1}^{(k)} = E_1^{(k)} - E_0^{(k)}$, and $f_{\Delta m_s=+1}^{(k)} = E_2^{(k)} - E_0^{(k)}$. Finally, the full set of eight resonance frequencies can be assembled from the pairs for each NV orientation:

$$f^{(Res)} = \{ f_{\Delta m_s=-1}^{(k=1)}, \quad f_{\Delta m_s=+1}^{(k=1)}, \quad f_{\Delta m_s=-1}^{(k=2)}, \quad f_{\Delta m_s=+1}^{(k=2)}, \quad f_{\Delta m_s=-1}^{(k=3)}, \quad f_{\Delta m_s=+1}^{(k=3)},$$
$$f_{\Delta m_s=-1}^{(k=4)}, \quad f_{\Delta m_s=+1}^{(k=4)} \}$$

This fit function can be compared to the set of eight measured resonance frequencies $f_j^{(Res)}$, allowing the sum of squares to be minimized with respect to the model parameters $\vec{B}$ and $D_{ZFS}^{(k)}$.

Assuming sufficient SNR in the spectral data, this fitting procedure generally does not result in incorrect parameter estimation due to local minima in the least-squares objective function. We ascertain this from the observed smoothness of the magnetic field maps obtained. The robustness of the fit is attributed to good *a-priori* knowledge of the B parameter(s) when the external bias field is large compared to the fields produced by the sample. When the bias field is comparable to or smaller than unknown sample fields or diamond strain, greater care must be taken in generating appropriate initial parameter guesses for the Levenberg-Marquardt algorithm. Failure to supply the algorithm with good initial parameters can be identified by non-convergence of fits, or by the appearance of added noise or patchiness in the resulting magnetic field maps.

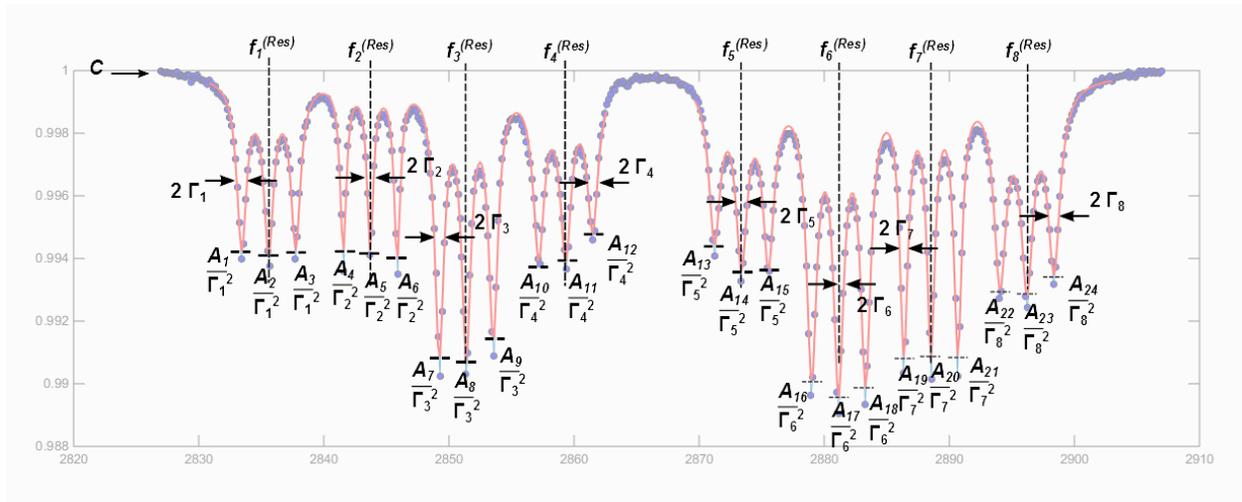

**Figure S4:** Typical VMM spectrum taken from a single magnetic image pixel (copied from Figure 1c of the main text). All VMM spectral fit parameters are labeled, including hyperfine amplitude parameters $A_{1...24}$, resonance frequencies $f_{1...8}^{(Res)}$, resonance linewidth parameters $\Gamma_{1...8}$, and fluorescence offset $C$. (A similarly labeled figure for PMM or CPMM would contain only 6 independent amplitude parameters, 2 resonance frequencies, 2 linewidth parameters, and one fluorescence offset.) Because of the definition of the spectral fit function, the amplitude of each resonance is given by $/\Gamma^2$, and the FWHM linewidth is $2\,\Gamma$. For VMM, the resonance frequencies are used as inputs to a second nonlinear fit. In the second fit, the model function $f^{(Res)}(\vec{B}, D_{ZFS}^{(k)})$ is a list of 8 expected resonance frequencies, calculated as a function of the local vector magnetic field $\vec{B}$, and a set of local axial strain parameters $D_{ZFS}^{(k)}$, for $k = 1...4$. Minimizing the sum of squared differences between the measured $f_j^{(Res)}$ and the model $f^{(Res)}(\vec{B}, D_{ZFS}^{(k)})$ allows all three components of the local vector magnetic field to be determined, along with the diamond strain parameters.

## Supplementary Figure S5: Physical Mechanism of Circularly Polarized Magnetic Microscopy (CPMM)

Because of its ability to operate at low bias fields ($B_0 < 10$ µT), circularly-polarized magnetic microscopy (CPMM) is expected to be the preferred mode of QDM operation for many geological applications. However, the physical mechanisms underlying this technique are somewhat complex for an ensemble of NV centers with orientations distributed along all four [111] diamond axes; we therefore provide here an extended discussion, starting with a single NV orientation before returning to the general case.

We begin with a qualitative picture of the rationale for CPMM measurements, with only one NV orientation. We consider NV centers formed from $^{15}$N, which we use for all CPMM measurements because their hyperfine structure is simpler than that of NV centers formed from $^{14}$N. The optically detected magnetic resonance (ODMR) spectrum for $^{15}$NV at zero applied magnetic field and zero strain, when excited with linearly-polarized microwaves, consists of two doubly-degenerate transitions. They are split by the $^{15}$N hyperfine interaction, which is the dominant interaction in the absence of applied fields or strain. The spectral line to the blue of the ZFS frequency consists of two overlapping transitions, one with $\Delta m_s = +1$ and $m_I = +½$, one with $\Delta m_s = -1$ and $m_I = -½$. The same is true for the line to the red of the ZFS, except one transition has $\Delta m_s = +1$ and $m_I = -½$, and the other $\Delta m_s = -1$ and $m_I = +½$. If a small magnetic field is applied parallel to the NV axis, the $\Delta m_s = +1$ transitions on both sides of the ZFS move slightly to higher frequency, and the $\Delta m_s = -1$ transitions both move slightly to lower frequency, resulting in an apparent broadening of the observed ODMR spectral lines (Figure S5a). If, on the other hand, a small magnetic field is applied anti-parallel to the NV axis, the $\Delta m_s = +1$ transitions both move slightly to lower frequency, and the $\Delta m_s = -1$ transitions both move slightly to higher frequency. The observed ODMR spectrum is the same as the first case.

This situation is disadvantageous for magnetic field sensing, for two reasons: (i) the sign of the applied field cannot be determined from the spectrum, and (ii) the magnitude of the applied field must be ascertained from the broadening of the spectral lines, rather than a shift in their mean, which is easier to measure experimentally. To break the symmetry between $\Delta m_s = +1$ and $\Delta m_s = -1$ transitions, we can drive the transitions with microwaves that are right circularly-polarized about the NV axis, which are perfectly selective for $\Delta m_s = +1$. Now, a magnetic field parallel to the NV axis causes both lines to shift to higher frequency, and a magnetic field anti-parallel to the NV axis causes both lines to shift to lower frequency. If we switch to left circularly-polarized microwaves, the opposite happens: we drive $\Delta m_s = -1$, with the result that a magnetic field parallel to the NV axis causes both lines to shift to lower frequency and a magnetic field anti-parallel to the NV axis causes both lines to shift to higher frequency (Figure S5b). By dithering between left- and right- circularly-polarized microwaves, we can determine both the sign and the magnitude of small applied magnetic fields with high precision. In the following paragraphs, we elaborate on this qualitative picture using an approximate NV Hamiltonian, and extend it to the experimentally relevant case of four simultaneously-probed NV orientations.

The Hamiltonian for the ground electronic state of a $^{15}$NV center (with electronic spin $S = 1$ and nuclear spin $I = 1/2$) may be written as follows [Felton *et. al.* 2009, Doherty *et. al.* 2013, Barson *et. al.* 2017]:

$$H = D_{ZFS} S_3^2 + \vec{S} \cdot \overleftrightarrow{M} \cdot \vec{S} + \vec{S} \cdot \overleftrightarrow{A} \cdot \vec{I} + \mu_B \vec{B} \cdot \overleftrightarrow{g} \cdot \vec{S} + g_N \mu_N \vec{B} \cdot \vec{I}$$

In this expression, which is valid in the absence of externally-applied electric fields, $\vec{S}$ and $\vec{I}$ are the electronic and nuclear spin operators; $\vec{B}$ is the local magnetic field; $D_{ZFS}$ is the NV zero field splitting (ZFS) energy, $\vec{M}$ is the strain interaction tensor, $\vec{A}$ is the electronic-nuclear hyperfine interaction, $\vec{g}$ is the electronic Zeeman tensor, and $g_N$ is the isotropic nuclear g-factor; and $\mu_B \approx 14$ GHz/T and $\mu_N \approx 7.6$ MHz/T are the Bohr magneton and nuclear magneton. The values of the interaction tensors ($\vec{M}, \vec{g}, \vec{d}$) have been measured in electron paramagnetic resonance and/or uniaxial stress experiments [Doherty *et. al.* 2013, Barson *et. al.* 2017].

For the sake of clarity, we adopt a simpler description, valid in the limit of small magnetic field, $\mu_B B \ll A_\parallel \ll D_{ZFS}$ (where $A_\parallel = d^{HF} = 3.03$ MHz is the axial component of the hyperfine tensor and $D_{ZFS} = 2.87$ GHz). Because the magnetic field is taken to be small ($B \ll \frac{A_\parallel}{\mu_B} \approx 200$ µT), we need to consider only the component $B_\parallel$ aligned along the NV axis; perpendicular components interact via the transverse spin operators $S_1$ and $S_2$, which are suppressed by the large ZFS and result in resonance line shifts only on the order of $\frac{(\mu_B B)^2}{D_{ZFS}} \approx 10^{-3} \mu_B B$. Then, noting that $\vec{g}$ is nearly isotropic [Felton *et. al.* 2009] and $\mu_N \ll \mu_B$, the Hamiltonian can be written as:

$$H \approx D_{ZFS} S_3^2 + d^{HF} S_3 I_3 + g \mu_B B_\parallel S_3$$

Now, we consider the energy levels of a single NV orientation aligned along $\hat{u}_{NV}^{(1)} = \left\{-\sqrt{\frac{2}{3}}, 0, \sqrt{\frac{1}{3}}\right\}$ in the diamond coordinate frame. We begin with $B = 0$, and diagonalize this Hamiltonian in the $|m_S, m_I\rangle$ basis. This yields three doubly degenerate subspaces: $\left|0, \pm\frac{1}{2}\right\rangle$ with energy $= 0$; $\left|\pm 1, \mp\frac{1}{2}\right\rangle$, with energy $E = D_{ZFS} - \frac{d^{HF}}{2}$; and $\left|\pm 1, \pm\frac{1}{2}\right\rangle$ with energy $E = D_{ZFS} + \frac{d^{HF}}{2}$. Using a linearly-polarized microwave drive, the resulting ODMR spectrum contains two degenerate doublets (Figure S4a). If we now add a small static magnetic field along $\hat{u}_{NV}^{(1)}$ (denoted $B_\parallel$), the states with $m_S = +1$ are shifted upwards by $g \mu_B B_\parallel$, and the states with $m_S = -1$ are shifted down by the same amount. For shifts smaller than the resonance linewidths (i.e., $B_\parallel \ll \frac{\Gamma}{g \mu_B} \approx 10$ µT, for typical linewidth parameters $\Gamma \sim 300 - 500$ kHz in our diamonds), the spin transitions for the two states are not resolved, and the ODMR spectrum under linear microwave drive appears broadened (Figure S5a).

Experimentally, determination of the magnetic field $B_\parallel$ from the broadened ODMR transition is difficult; we would prefer a signal that is proportional to the mean, rather than the second moment, of the resonance line shape. We can accomplish this by driving the ODMR transitions with microwaves that are circularly-polarized with respect to $\hat{u}_{NV}^{(1)}$. Using a right-circular (σ⁺) polarization selectively drives only transitions to the $m_S = +1$ state, which shifts to higher frequency for $B_\parallel > 0$, and lower frequency for $B_\parallel < 0$. Left-circular (σ⁻) polarization drives transitions to $m_S = -1$, which are shifted with the opposite sign. By alternating between σ⁺ and σ⁻ drives, and subtracting the fitted resonance frequencies for the two transitions, we can efficiently estimate the magnitude and sign of $B_\parallel$ (Figure S5b).

Some modification is needed to generalize this approach to an ensemble of NV centers equally distributed over all four possible orientations, $\hat{u}_{NV}^{(k)}$, $k = 1\ldots4$. First, we note that it is not possible for the microwave drive to be circularly polarized along all orientation directions simultaneously. However, there are three unit vectors in our global *x,y,z* coordinate system [namely, $\hat{v}_1 = \frac{1}{\sqrt{2}}(\hat{x} + \hat{y})$, $\hat{v}_2 = \frac{1}{\sqrt{2}}(\hat{x} - \hat{y})$, and $\hat{v}_3 = \hat{z}$] that have equal projections onto all NV axes, satisfying $\hat{v}_j \cdot \hat{u}_{NV}^{(k)} = \sqrt{\frac{1}{3}}$ for all *k*. A circularly-polarized drive along one of these vectors will resolve into the sum of linear and circular components along each of the NV orientations $\hat{u}_{NV}^{(k)}$, with equal circular component for each orientation. Suppose a static magnetic field is applied along one particular NV orientation, say $\vec{B}^{(1)} = B\,\hat{u}_{NV}^{(1)}$, with the microwave drive circularly-polarized about one of the equal-projection axes. Ignoring NV orientations $\hat{u}_{NV}^{(k=2-4)}$ for the moment, we realize that the transitions for NV centers oriented along $\hat{u}_{NV}^{(1)}$ will be shifted exactly as before, except that the linear portion of the microwave drive will excite transitions to both $m_S = +1$ and $m_S = -1$, reducing the signal contrast and necessitating a more complicated fit. In practice, to facilitate straightforward alignment of the experiment, we typically choose $\hat{v}_3 = \hat{z}$ as our circular polarization axis (Figure S5c-d).

The applied magnetic field $\vec{B}^{(1)}$ will, of course, also cause shifts in the transition frequencies of NV centers in the other 3 orientations, and the observed ODMR spectrum will in general be a superposition of fluorescence from all 4 orientations. We would now like to understand the shape of this superposed ODMR spectrum for a general applied field $\vec{B}$ that is not necessarily aligned along any NV axis. Because the frequency shifts for each orientation depend, to lowest order, only on the component of $\vec{B}$ parallel to the NV axis, it turns out that the overall shift observed in the full ODMR spectrum is proportional only to the component of the applied $\vec{B}$ field that is parallel to the circular polarization axis $\hat{v}_3 = \hat{z}$. To see this, note that according to the definitions in Supplementary Figure S3, $\hat{u}_{NV}^{(1)} \cdot \hat{x} = -\hat{u}_{NV}^{(2)} \cdot \hat{x}$, with $\hat{u}_{NV}^{(3,4)} \cdot \hat{x} = 0$, and $\hat{u}_{NV}^{(3)} \cdot \hat{y} = -\hat{u}_{NV}^{(4)} \cdot \hat{y}$, with $\hat{u}_{NV}^{(1,2)} \cdot \hat{y} = 0$. Thus, any component of $\vec{B}$ in the x-y plane will cause half of the NV centers to experience positive frequency shifts and half negative, of equal magnitude, irrespective of the handedness of the applied circular polarization. The resulting superposed ODMR spectrum will show an apparent broadening of the respective lines, but no change in their mean transition frequency. On the other hand, $\hat{u}_{NV}^{(1)} \cdot \hat{z} = \hat{u}_{NV}^{(2)} \cdot \hat{z} = \hat{u}_{NV}^{(3)} \cdot \hat{z} = \hat{u}_{NV}^{(4)} \cdot \hat{z} = \sqrt{\frac{1}{3}}$. This means that the component of $\vec{B}$ in +z-direction produces a positive shift in the energy of the $m_S = +1$ state and a negative shift in the energy of the $m_S = -1$ state for all NV orientations. Changing between σ⁺ and σ⁻ microwave drive polarizations will respectively select for the transitions to $m_S = +1$ or $m_S = -1$, (imperfectly, because the projected polarization along each NV axis is elliptical), and the resulting ODMR lines will be shifted up and down by $\Delta f = \pm g\,\mu_B\,\frac{B_z}{\sqrt{3}}$. Thus, our CPMM measurements are sensitive only the z-component of the applied magnetic field.

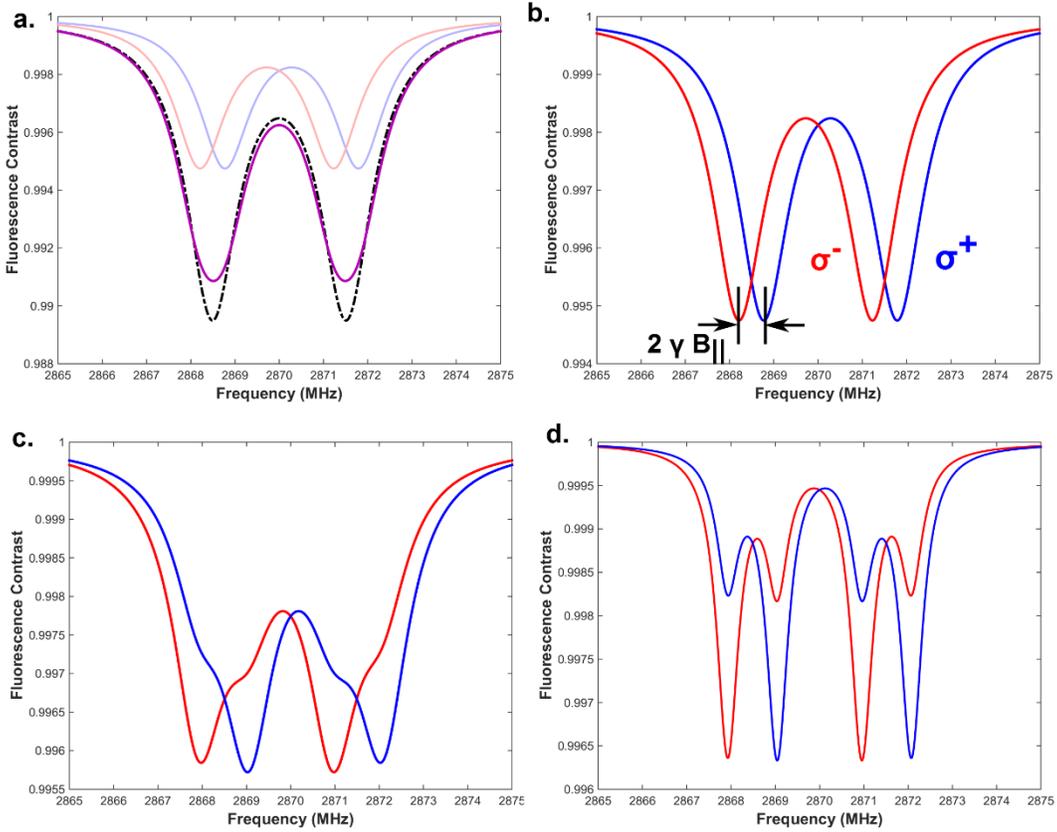

**Figure S5:** Simulated $^{15}$NV CPMM spectra for NV centers of a single orientation, $\hat{u}_{NV}^{(1)}$, under magnetic field $B_{||}$ applied parallel to the NV axis. (In panels (a)-(c), amplitude and linewidth parameters are A = 0.005 and $\Gamma$ = 700 kHz, using the same notation as Figure S4.) **(a)** Black dashed line shows ODMR for the case of $B_{||}$=0, resulting in a pair of doubly-degenerate lines separated by the hyperfine splitting $d^{HF}$ = 3.03 MHz. Adding $B_{||}$ = 10 µT lifts the degeneracy between the $m_s$=0→+1 (blue line) and the $m_s$=0→-1 (red line) transitions. However, under linearly-polarized microwave drive, these cannot be observed separately; only their sum (purple line) can be detected, resulting in a slightly broadened ODMR feature compared to $B_{||}$ = 0. **(b)** For $B_{||}$ =10 µT, driving with microwaves that are $\sigma^+$-polarized (blue line) or $\sigma^-$-polarized (red line) with respect to $\hat{u}_{NV}^{(1)}$ gives perfect distinguishability of the $m_s$ transitions. $B_{||}$ is determined from the shift in the center frequency of the ODMR lines under $\sigma^+$ and $\sigma^-$ driving. **(c)** ODMR spectrum when the circular polarization axis is $\hat{z}$, instead of $\hat{u}_{NV}^{(1)}$, resulting in reduced suppression of both the $m_s$=0→-1 transition under $\sigma^+$ driving (blue line) and the $m_s$=0→+1 transition under $\sigma^-$ driving (red line). However, the different $m_s$ transitions can still be distinguished and fit to determine $B_{||}$. Here, we used $B_{||}$ = 20 µT to make the Zeeman splitting easily visible. **(d)** Same as (c), but now using linewidth $\Gamma$ = 300 kHz. With these parameters, the simulated spectrum qualitatively resembles the experimental data in Figure 1e of the main text.

# Supplementary Figure S6: Full Vector Data for ALHA81001

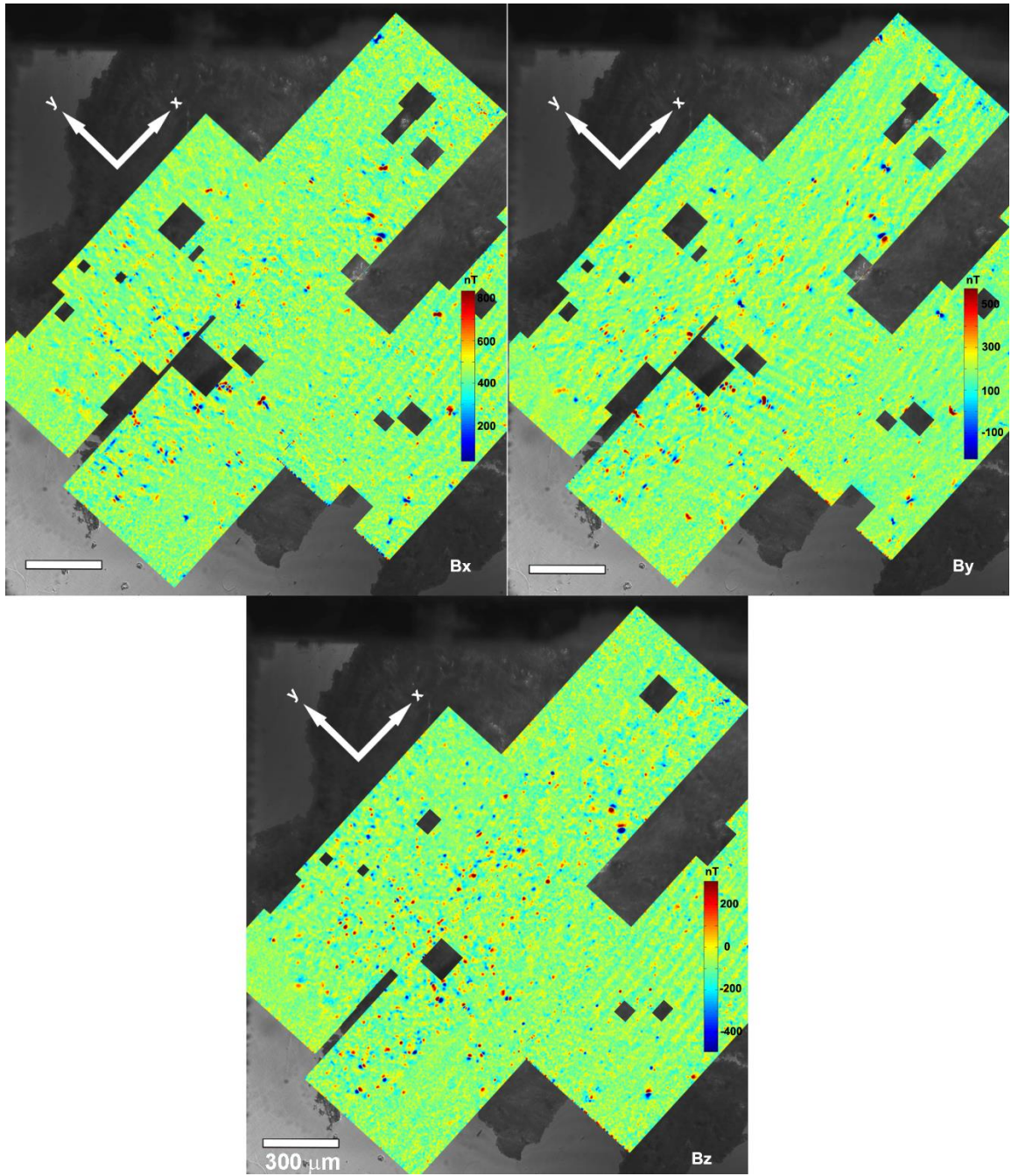

**Figure S6:** Full vector (three-axis) magnetic maps for a section of ALHA81001, acquired in VMM using diamond D1. The $B_z$ image is repeated from Fig. 2d of the main text. Field maps for $B_x$ and $B_y$ were constructed from the same data set as $B_z$, using the fitting procedure described in Supplementary Figure S4. The nonzero mean magnetic field in each map is due to finite precision in the reversal of $B_0$, the magnetic bias field. Blank squares in the maps correspond to areas that were manually removed due to the presence of known strain defects in the diamond sensor. These areas are larger in $B_x$ and $B_y$, due to the stronger effect of strain on the vales obtained for those field components using our fitting procedure.

# Supplementary Figure S7: ALHA81001 $B_z$ Images With Two Diamond Chip Orientations

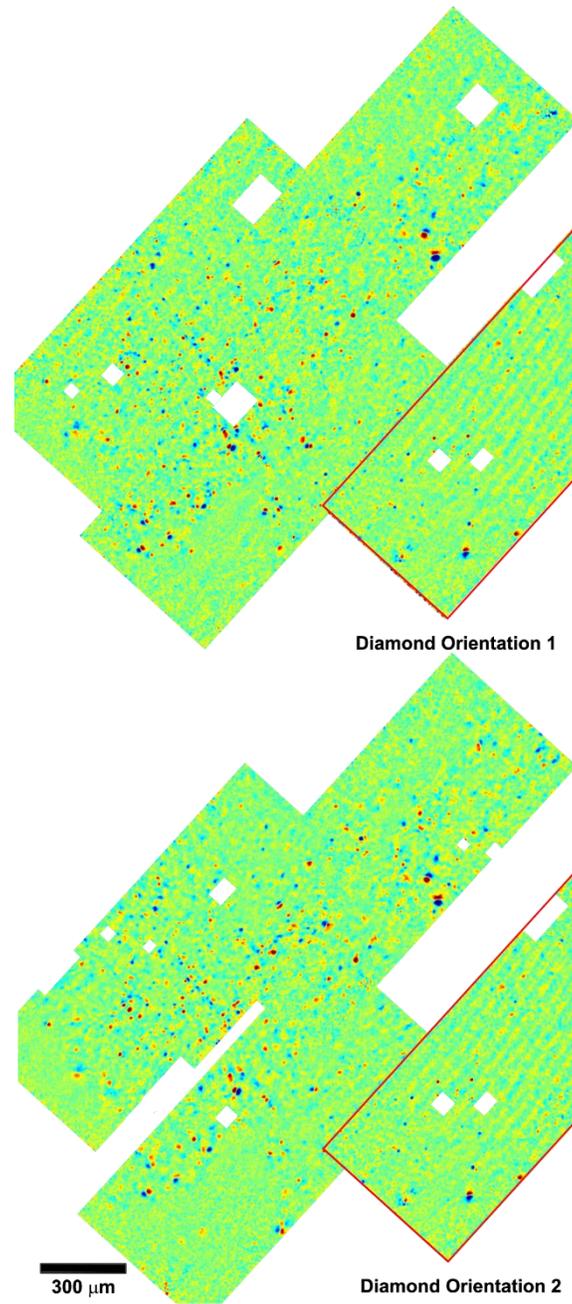

**Figure S7:** Magnetic maps for a 30 μm thin section of ALHA81001, acquired in VMM using diamond D1. Only the out-of-plane ($B_z$) component is displayed. Maps for each of the tiled FOVs were acquired twice, with the diamond chip physically rotated 90° between acquisitions. (Exception: The fourth FOV, outlined in red, was acquired only for one orientation. It is nevertheless included here for straightforward comparison to Fig. 2d of the main text.) Magnetic features that appear in maps obtained with both diamond orientations are assumed to be produced by real sources in the rock sample; features that appear for only one diamond orientation are likely due to localized strain defects or diamond surface contamination.

## Supplementary Figure S8: Sensitivity Scaling

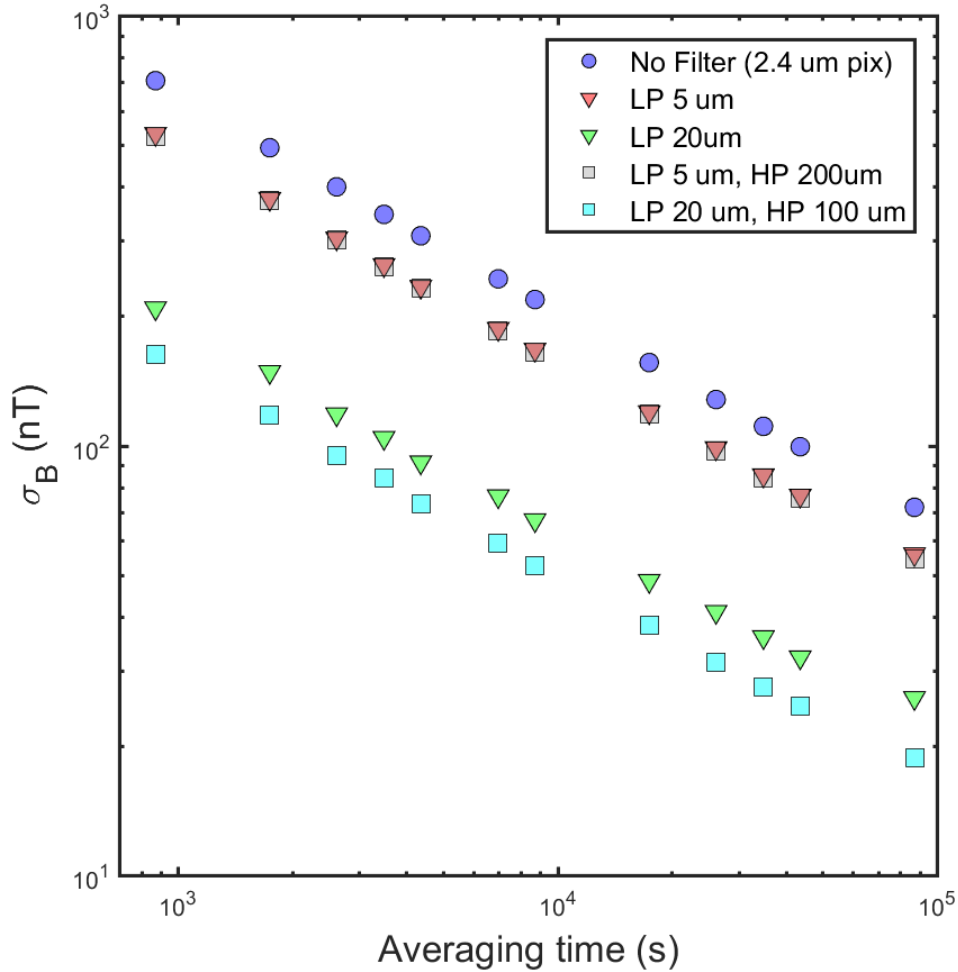

**Figure S8:** Averaging properties of QDM magnetic field sensitivity. Data were acquired in VMM mode, with an empty (magnetically uniform, with top surface of the diamond in air) FOV of 1.0 mm × 0.5 mm, using diamond D4. Imaging optics were chosen such that each un-binned image pixel corresponds to a (2.4 µm)$^2$ region of the sensor. Averaged ODMR measurements were continuously acquired at a rate of 44 seconds per average, with each average consisting of $Q$ = 500 camera exposures at uniformly-spaced frequencies $f_q$ (with $q$ = 1…$Q$) of the spin drive field $B_1$. After a given averaging time $T_{avg}$, the data were fit using the standard procedure for VMM as described in Supplementary Figure S3. The map of the $B_z$ component of the extracted magnetic field was then spatially filtered using several distinct combinations of high-pass (Butterworth, third order, applied in the frequency domain) and low-pass (Gaussian, applied by convolution in real-space domain) filters. The standard deviation over all image pixels was then calculated for the $B_z$ component of the filtered, fitted field maps and plotted above. Noise scaling for all spatial filter combinations was nearly ~$T_{avg}^{-½}$ up to $10^5$ seconds. The image-area-normalized sensitivity represented by the grey squares (and the red triangles) is ~75 µT·µm ·Hz$^{-½}$, which is comparable (within a factor of ~3) to the demonstrated sensitivity for the ALH81001 magnetic images shown in Figure 2d of the main text, using here a somewhat lower optical excitation intensity and the same post-processing procedure.

## Supplementary Figure S9: Energy Dispersive Spectroscopy on ALHA81001

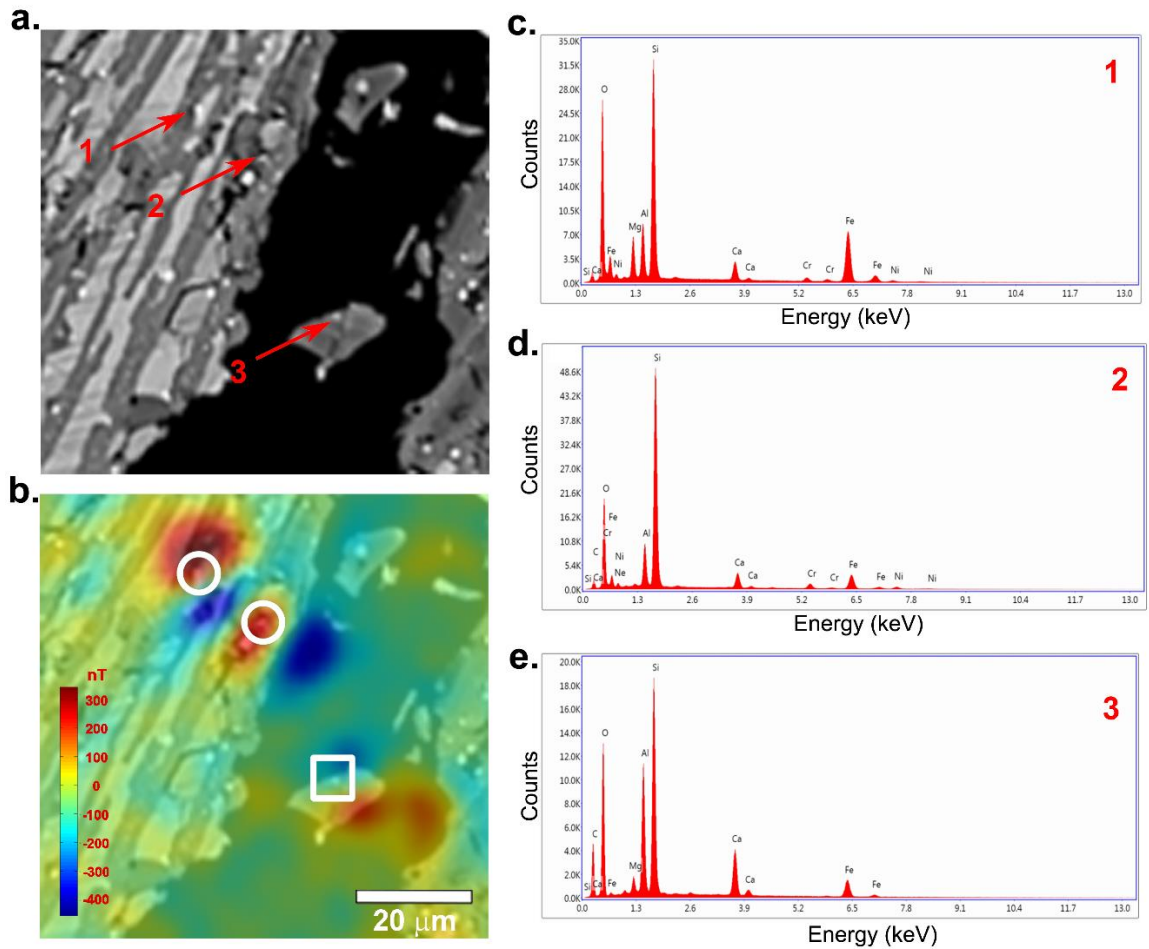

**Figure S9:** Combined QDM and qualitative compositional analysis of ALHA81001. Subfield shown is same as inset in Main Text Fig. 2. **(a)** Backscattered electron (BSE) map of high magnetization region in ALHA81001. High atomic number grains corresponding to strong magnetic sources are labeled in red. **(b)** QDM map of the same field of view as part (a), acquired in VMM using diamond D1. **(c-e)** Energy dispersive spectroscopy (EDS) spectra of grains 1 through 3, respectively, as labeled in part (a). The <1 micrometer size of the grains preclude isolation of their EDS spectra from those of the surrounding pyroxene and plagioclase phases, resulting in strong Si, Al, Ca, and Mg peaks. The presence of Fe, Cr, and Ni in grains 1 and 2 suggests a chromite composition. The spectrum of grain 3 suggests a low-Ni Fe metal phase, likely kamacite. The Curie temperatures of chromite and kamacite are consistent with previous paleomagnetic results on ALHA81001 (Main Text). Cohenite is an alternative interpretation for grain 3, although the carbon peak may be due to uneven carbon deposition during sample preparation. Other high atomic number phases near the magnetic sources consist of troilite and ilmenite, which are not ferromagnetic at room temperature.

## Supplementary Figure S10: Effect of NV Layer Thickness on QDM Spatial Resolution

The NV layer thickness, $t_{NV}$, is an important parameter in QDM sensor design. QDM magnetic field sensitivity varies as the square root of the per-pixel fluorescence intensity (assuming shot-noise limited detection); and this intensity is linearly proportional to the number of NVs probed. Thus, a diamond with an NV layer of thickness $t_{NV}$ = 10 µm provides a sensitivity improvement of ~30 compared to a diamond with $t_{NV}$ = 10 nm, for the same pixel area. However, magnetic fields produced by dipolar sources placed close to the sensor (i.e., $d_{s-s} \lesssim t_{NV}$ for $d_{s-s}$ the distance between the source and the diamond surface) will vary significantly over the sensing volume, potentially resulting in a loss of spatial resolution and/or a decrease in the fluorescence-averaged ODMR line shift in a single pixel.

We investigate this tradeoff numerically, using the geometry shown in Figure S10a. A dipolar source with magnetic moment $\vec{m} = m_0 \, \hat{m}$ is placed a distance $d_{s-s}$ from the diamond, which has NV layer thickness $t_{NV}$. The magnetic field at a distance $R$ from the dipole, projected onto a single NV axis $\hat{u}_{NV}$, is given by

$$\hat{u}_{NV} \cdot \vec{B}(\vec{R}) = \frac{\mu_0 m_0}{4\pi} [(3 \, (\hat{u}_{NV} \cdot \hat{R}) (\hat{m} \cdot \hat{R}) - \hat{u}_{NV} \cdot \hat{m})] \frac{1}{R^3} .$$

We can decompose $R$ into radial ($r$) and axial ($z$) components, $R^2 = (d_{s-s} + z)^2 + r^2$, and calculate the $z$-integrated fluorescence signal as a function of $r$ to determine the transverse spatial resolution of the sensor. For simplicity, we consider only a single Lorentzian line shape for the NV ODMR spectral response function, although the calculation generalizes straightforwardly to more realistic ODMR spectra. The $z$-integrated fluorescence signal $S$, as a function of $r$ and the ODMR frequency $f$, is given by

$$\begin{aligned} S(r, f) &= \int_{d_{s-s}}^{d_{s-s}+t_{NV}} \frac{A}{[f - g \, \mu_B \, B(R)]^2 + \Gamma_{NV}^2} dz \ . \\ &= \int_{d_{s-s}}^{d_{s-s}+t_{NV}} \frac{A}{\left[f - \frac{g \, \mu_B \, B_s \, d_{s-s}^3}{(r^2+z^2)^{3/2}}\right]^2 + \Gamma_{NV}^2} dz , \end{aligned}$$

for $g \approx 2$ the Landé g-factor, $\mu_B \approx 14$ GHz/T the Bohr magneton, and $B_s = \frac{\mu_0 \, m_0}{4\pi \, d_{s-s}^3} [3 \, (\hat{u}_{NV} \cdot \hat{R}) (\hat{m} \cdot \hat{R}) - \hat{u}_{NV} \cdot \hat{m}]$ the field produced by the dipole at the top surface of the diamond. Here, we have defined the Lorentzian line shape parameters such that $f$ = 0 is the center frequency when no magnetic field is applied, $\Gamma_{NV}$ is the half-width at half-max (HWHM) linewidth parameter, and $A / \Gamma_{NV}^2$ is the peak ODMR contrast that would be observed for a vanishingly-thin NV layer.

We now change to dimensionless (normalized) coordinates by making the following substitutions:

$$\rho = r/d_{s-s}, \quad \xi = z/d_{s-s}, \quad \tau = t_{NV}/d_{s-s}, \quad \phi = f/\Gamma_{NV}, \quad S_0 = \frac{A \, d_{s-s}}{\Gamma_{NV}^2}, \quad \beta_s = \frac{g \, \mu_B \, B_s}{\Gamma_{NV}}$$

The integrated fluorescence becomes

$$S(\rho, \phi) = S_0 \int_1^{1+\tau} \frac{1}{\left[\phi - \frac{\beta_s}{(\rho^2+\xi^2)^{3/2}}\right]^2 + 1} d\xi \ .$$

To extract the measured magnetic field $\beta_{meas}$ at each position $\rho$ from the fluorescence signal, we would normally fit $S$ to a Lorentzian and find the line center. In the present calculation, however, to avoid complications associated with asymmetric broadening of the integrated fluorescence line shape when the magnetic source is very close to the sensor (i.e., when $\tau \gtrsim 1$), we simply take the frequency of the line peak, $\phi_{pk}$, to represent the average value of line shift caused by the magnetic field in the sensor (Figure S10b):

$$\beta_{meas}(\rho) = \phi_{pk}(\rho) = \underset{\phi}{\mathrm{argmax}}[S(\rho,\phi)]$$

We now plot the observed magnetic field as a function of $\rho$ for various experimentally relevant values of the normalized NV layer thickness $\tau$ and the normalized magnetic field at the diamond surface $\beta_s$ (Figure S10c–f). The minimum achievable standoff distance is on the order of $d_{s-s} \approx 1$ µm due to sample roughness, and the NV layer thickness $t_{NV}$ can range from 10 nm to 10 µm, giving normalized layer thickness $\tau \sim 0.01 - 10$. (We could also choose larger standoff distances by using spacers between the diamond and the sample, in which case $\tau$ becomes even smaller and the QDM approaches an ideal 2D detector.) Typical values for the NV linewidth parameter are on the order of $\Gamma_{NV} \approx 500$ kHz, and the characteristic scale for magnetic fields produced by most rock samples we have studied is $B_s \approx 100$ nT – 100 µT. This gives normalized surface fields $\beta_s \sim 0.01 - 10$.

Two broad conclusions may be drawn from the plots of $\phi_{pk}(\rho)$ shown in Figures S10c – f:
1. The transverse spatial resolution of the QDM does not depend strongly on the NV layer thickness. For nearly all values of $\tau$, the measured ODMR line shift decreases to half of its maximum value at $\rho \sim 1 - 1.5$. This is because the line shift in the z-integrated fluorescence is dominated by NV centers near the surface of the sensor, with deeper NV centers (which experience a weaker magnetic field) contributing mainly to a broadening of the line. In effect, the deep NV centers add little to the QDM sensitivity for large $\tau$, but also do not significantly harm the transverse spatial resolution. This is true for both weak ($\beta_s < 1$) and strong ($\beta_s > 1$) magnetic sources.
2. For NV layers of finite thickness ($\tau \gg 0$), the maximum ODMR line shift $[\phi_{pk}(\rho = 0)]$ measured from the z-integrated fluorescence is significantly smaller than the corresponding value for an infinitesimal surface layer. This effect is most pronounced with strong magnetic sources ($\beta_s > 1$), for which deep NV centers contribute appreciably to the signal to produce significant line-broadening. In order to extract an accurate value for the source magnetic moment $m_0$ from the measured line shift, a $\tau$-dependent scale factor must be applied to the measured field map unless $\tau \ll 1$. The range of values of $d_{s-s}$ in a measurement of a geological sample containing multiple dipole sources (denoted $\Delta d_{s-s}$) is set by the thickness of the sample: $\Delta d_{s-s} = t_{samp}$. When measuring using a diamond sensor of appreciable thickness (i.e., $t_{NV} \sim \overline{d_{s-s}}$, for $\overline{d_{s-s}}$ the distance from the diamond surface to the middle of the geological sample in z) it is important to prepare thin samples, $t_{samp} \ll \overline{d_{s-s}}$, This ensures that the distance between the magnetic sources and the diamond surface is approximately constant, and the scale factor is well-defined. In practice, this may be technically challenging, since polishing samples too thin may result in shape anisotropy or magnetostriction effects.

We note that these considerations are not relevant to the magnetic field maps shown in Figures 1 and 2 of the main text, which were acquired with thin-NV-layer diamonds D1 and D2, with $\tau \lesssim 0.01$. The data of Figure 4 were acquired with diamond D3, with $\tau \sim 0.1 - 1$, and thus might be subject to finite layer-thickness corrections. However, in this case, the goal of the measurement was localization of sources, rather than quantitative determination of their magnetic moments.

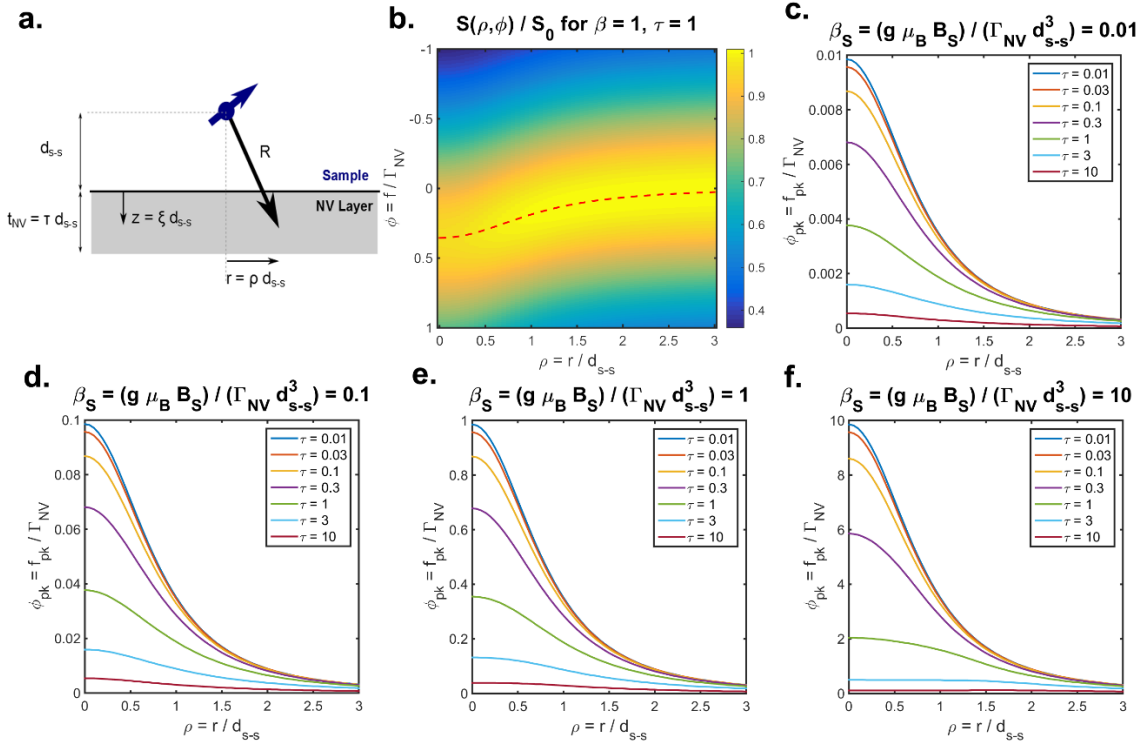

**Figure S10: (a)** Schematic showing geometry of our model for the z-integrated ODMR signal. A dipolar source is placed a distance $d_{s\text{-}s}$ above the diamond. The ODMR spectrum is calculated at every point ($r,z$) in the NV layer, then integrated with respect to $z$ to obtain the observed magnetic field as a function of $r$. Normalized coordinates $\tau$, $\xi$, and $\rho$ replace $t_{NV}$, $z$ and $r$, respectively. **(b)** Characteristic plot of the z-integrated ODMR signal $S(\rho,\phi)$, for normalized surface magnetic field $\beta_s = 1$ and normalized NV layer thickness $\tau = 1$. Each vertical slice is an integrated ODMR spectrum (approximately Lorentzian). The red dashed line traces the peak of the ODMR spectra, $\phi_{pk}(\rho)$, which is taken as the best estimate of the measured magnetic field in the NV layer as a function of $\rho$. **(c – f)** Plots of $\phi_{pk}(\rho)$, for experimentally relevant normalized layer thicknesses $\tau = 0.01 - 10$. Each panel is calculated for a different dipolar source strength, parameterized by the normalized magnetic field it produces at the diamond surface, $\beta_s = 0.01 - 10$. For an infinitesimally thin NV layer, we expect $\phi_{pk}(\rho = 0) = \beta_s$.

# Supplementary Figure 11: Expected $B_z$ Calculation for QDM Calibration

To calibrate the absolute accuracy of magnetic field measurements using the QDM in various modes of operation, we applied homogeneous static fields to the diamond chip using several well-characterized solenoids. Each solenoid was constructed with ten turns of wire, the precise positions of which were measured in an optical microscope. We calculated the expected field at the diamond sensor, located at a known position along the solenoid axis, using the following expression for the magnetic field above a given loop at distance $h$ from the NV sensing layer:

$$B_{loop} = \frac{\mu_0}{4\pi} \frac{2\pi a^2 I}{(a^2+h^2)^{3/2}}.$$

Here, $B_{loop}$ is the magnetic field along the loop axis, $\mu_0 = 4\pi \times 10^{-7}$ T·m/A is the vacuum permeability, $a$ is the loop radius, and $I$ is the current through the loop. To estimate the total field from ten loops, we calculated $B_{loop}$ for the loops (each with a different $h$) and summed them.

As an example, Fig. S11 shows a photo of one calibration solenoid (used for the PMM calibration at 18.6 mT in Fig. 3b of the main text). We measured the spacing of the solenoid wire loops ($\Delta h = 0.48 \pm 0.02$ mm) and the separation between the nearest loop and the NV layer ($h_0 = 20.9 \pm 0.1$ mm) with an optical microscope. Using a micrometer, we measured the loop radius $a = 15.5 \pm 0.05$ mm. This yields a total expected magnetic field along the loop axis

$$B_{total} = \frac{\mu_0}{4\pi} \sum_{n=0}^{9} \frac{2\pi a^2 I}{\left(a^2 + (h_0 + n\Delta h)^2\right)^{3/2}} = \left(71.0 \pm 0.8 \frac{\text{nT}}{\text{mA}}\right) \times I.$$

The applied magnetic field was along the z-axis, perpendicular to the diamond surface, whereas the field measured in a PMM experiment is the component projected along one NV axis. We therefore multiply the above expression by $\hat{u}_{NV}^{(1)} \cdot \hat{z} = 1/\sqrt{3}$ to yield 41.0 nT/mA as the expected projected field component due to the solenoid.

We used a laser current supply (similar to those used to drive the Helmholtz coils) to drive the calibration loop current, monitoring the current by measuring the voltage drop across a known resistance in series. The applied field was thereby known to an accuracy of ~0.6%. The uncertainty in the expected calibration slope was determined from uncertainties in the measured dimensions $h_0$, $\Delta h$, and $a$; all uncertainty estimates quoted here are 1 σ.

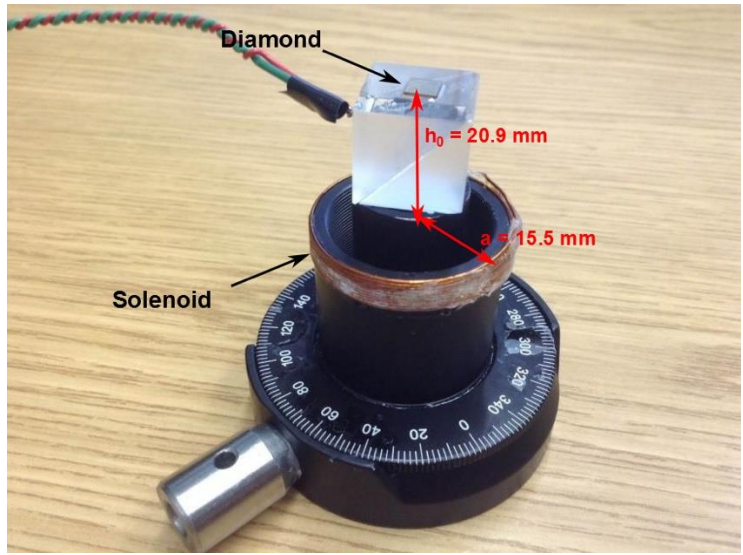

**Figure S11:** Photograph of the solenoid used for absolute magnetic field calibration of PMM measurements.